\documentclass[%
aip,
rsi,
 amsmath,amssymb,
 reprint,%
]{revtex4-1}

\usepackage{graphicx}% Include figure files
\usepackage{dcolumn}% Align table columns on decimal point
\usepackage{bm}% bold math
\usepackage[utf8]{inputenc}
\usepackage[T1]{fontenc}
\usepackage{mathptmx}
\usepackage{etoolbox}

\makeatletter
\def\@email#1#2{%
 \endgroup
 \patchcmd{\titleblock@produce}
  {\frontmatter@RRAPformat}
  {\frontmatter@RRAPformat{\produce@RRAP{*#1\href{mailto:#2}{#2}}}\frontmatter@RRAPformat}
  {}{}
}%
\makeatother
\begin{document}

\preprint{AIP/123-QED}

\title[]{Time-resolved ARPES with probe energy of 6.0/7.2 eV and switchable resolution configuration}
% Force line breaks with \\
\author{Mojun Pan}
\affiliation{ 
Beijing National Laboratory for Condensed Matter Physics and Institute of
Physics, Chinese Academy of Sciences, Beijing 100190, China
}
\affiliation{University of Chinese Academy of Sciences, Beijing 100049, China}

\author{Junde Liu}
\affiliation{ 
Beijing National Laboratory for Condensed Matter Physics and Institute of
Physics, Chinese Academy of Sciences, Beijing 100190, China
}
\affiliation{University of Chinese Academy of Sciences, Beijing 100049, China}

\author{Famin Chen}
\affiliation{ 
Southern University of Science and Technology, Shenzhen 518055, China
}

\author{Ji Wang}
\affiliation{ 
Beijing National Laboratory for Condensed Matter Physics and Institute of
Physics, Chinese Academy of Sciences, Beijing 100190, China
}
\affiliation{ 
Songshan Lake Materials Laboratory, Dongguan 523808, China
}

\author{ChenXia Yun}
\altaffiliation[Authors to whom correspondence should be addressed: ]{cxyun@iphy.ac.cn}
\affiliation{ 
Beijing National Laboratory for Condensed Matter Physics and Institute of
Physics, Chinese Academy of Sciences, Beijing 100190, China
}
\affiliation{University of Chinese Academy of Sciences, Beijing 100049, China}

\author{Tian Qian}
\altaffiliation[Authors to whom correspondence should be addressed: ]{tqian@iphy.ac.cn}
\affiliation{ 
Beijing National Laboratory for Condensed Matter Physics and Institute of
Physics, Chinese Academy of Sciences, Beijing 100190, China
}
\affiliation{University of Chinese Academy of Sciences, Beijing 100049, China}

\date{\today}% It is always \today, today,

\begin{abstract}
We present a detailed exposition of the design for time- and angle-resolved photoemission spectroscopy (tr-ARPES) using a UV laser source based on non-linear optical effect in ${\beta}$-$\rm{BaB_2O_4}$ (BBO) and $\rm{KBe_2BO_3F_2}$ (KBBF) crystals. The photon energy of the probe laser can be switched between 6.0 and 7.2 eV, with the flexibility to operate each photon energy setting under two distinct resolution configurations. Under the fully optimized energy resolution configuration, we achieve an energy resolution of 8.5 meV for 6.0 eV and 10 meV for 7.2 eV. Alternatively, switching to the other configuration enhances temporal resolutions, yielding a time resolution of 72 fs for 6.0 eV and 185 fs for 7.2 eV. To validate the performance and reliability of our setup, we conducted measurements on two typical materials: the topological insulator $\rm{MnBi_2Te_4}$ and the excitonic insulator candidate $\rm{Ta_2NiSe_5}$.
\end{abstract}

\maketitle
\section{\label{sec:level1}INTRODUCTION\protect\\}

Time and angle-resolved photoemission spectroscopy (tr-ARPES) is a cutting-edge technique that has been developed in the last decade. By combining ultrafast laser techniques with angle-resolved photoemission spectroscopy, tr-ARPES provides a unique capability to investigate non-equilibrium band structures and photo-induced phase transitions in solid-state systems. Its remarkable capabilities have generated significant interest among researchers and have been extensively applied in the investigation of various quantum materials, including topological materials \cite{wang2012measurement,sobota2012ultrafast,sanchez2016ultrafast,ciocys2020manipulating,sobota2014ultrafast}, high-$\rm{T_c}$ superconductors \cite{damascelli2003angle,perfetti2007ultrafast}, excitonic insulators \cite{tang2020non,okazaki2018photo}, charge-density-wave systems \cite{schmitt2008transient,petersen2011clocking,rohwer2011collapse}, Floquet band engineering \cite{wang2013observation,mahmood2016selective,zhou2023pseudospin}, and others\cite{cavalieri2007attosecond,tao2016direct}. 

Over the past few decades, several tr-ARPES instruments with diverse designs and configurations have been developed \cite{yang2019time,bao2022ultrafast,ishida2016high,chen2023time,sie2019time,rohde2016time,puppin2019time,mills2019cavity}. The majority of these instruments aim to develop suitable probe laser sources through the utilization of high harmonics generation (HHG) in noble gases\cite{chen2023time,sie2019time,rohde2016time,puppin2019time,mills2019cavity} or second-order nonlinear optical effects in optical crystals \cite{yang2019time,bao2022ultrafast,ishida2016high}. The probe beam generated by HHG offers advantages such as higher photon energy (>20 eV), a wide range of tunable wavelengths \cite{wang2023high}, and enhanced temporal resolution. On the other hand, instruments that employ optical nonlinear crystals for probe laser generation are more widely accessible and generally provide superior energy resolution.

Fourth harmonic generation (FHG) in ${\beta}$-$\rm{BaB_2O_4}$ (BBO) is a widely utilized technique for generating probe lasers, resulting in a photon energy of approximately 6.0 eV. By coupling the $\rm{KBe_2BO_3F_2}$ (KBBF) crystal \cite{wu1996linear,chen2009deep} with the prism technique, ultraviolet (UV) lasers with higher photon energy can be generated. Yang et al.\cite{yang2019time} achieved a probe photon energy of 6.7 eV with a temporal resolution of approximately 1 ps, while Bao et al\cite{bao2022ultrafast} achieved a tunable probe photon energy range of 5.9-7.0 eV, with the temporal resolution improved to 320 fs at 7.0 eV. 

In this study, we have developed a UV source with a photon energy of 7.2 eV and have achieved a fully optimized resolution of 185 fs. Notably, we have also incorporated a versatile design that facilitates convenient switching between 6.0/7.2 eV and different resolution configurations. This setup can be operated with energy resolutions of 8.7 meV for 6.0 eV and 10 meV for 7.2 eV, which is particularly advantageous for accurately probing fine features of electronic band structures. Furthermore, it can be easily reconfigured to achieve improved time resolutions of 72 fs for 6.0 eV and 185 fs for 7.2 eV, which is highly beneficial for investigating transient nonequilibrium processes. The combination of a two-color UV laser source with our compatibility design substantially enhances experimental efficiency. To demonstrate the performance of our setup, we conducted measurements on several materials including topological insulator $\rm{Bi_2Se_3}$ \cite{zhang2009topological,xia2009observation}, $\rm{MnBi_2Te_4}$ \cite{li2019dirac} and an excitonic insulator candidate $\rm{Ta_2NiSe_5}$ \cite{Wakisaka2009, Mor2017, Fukutani2021}.

\begin{figure*}[ht]
    \centering
    \includegraphics[width=1\textwidth]{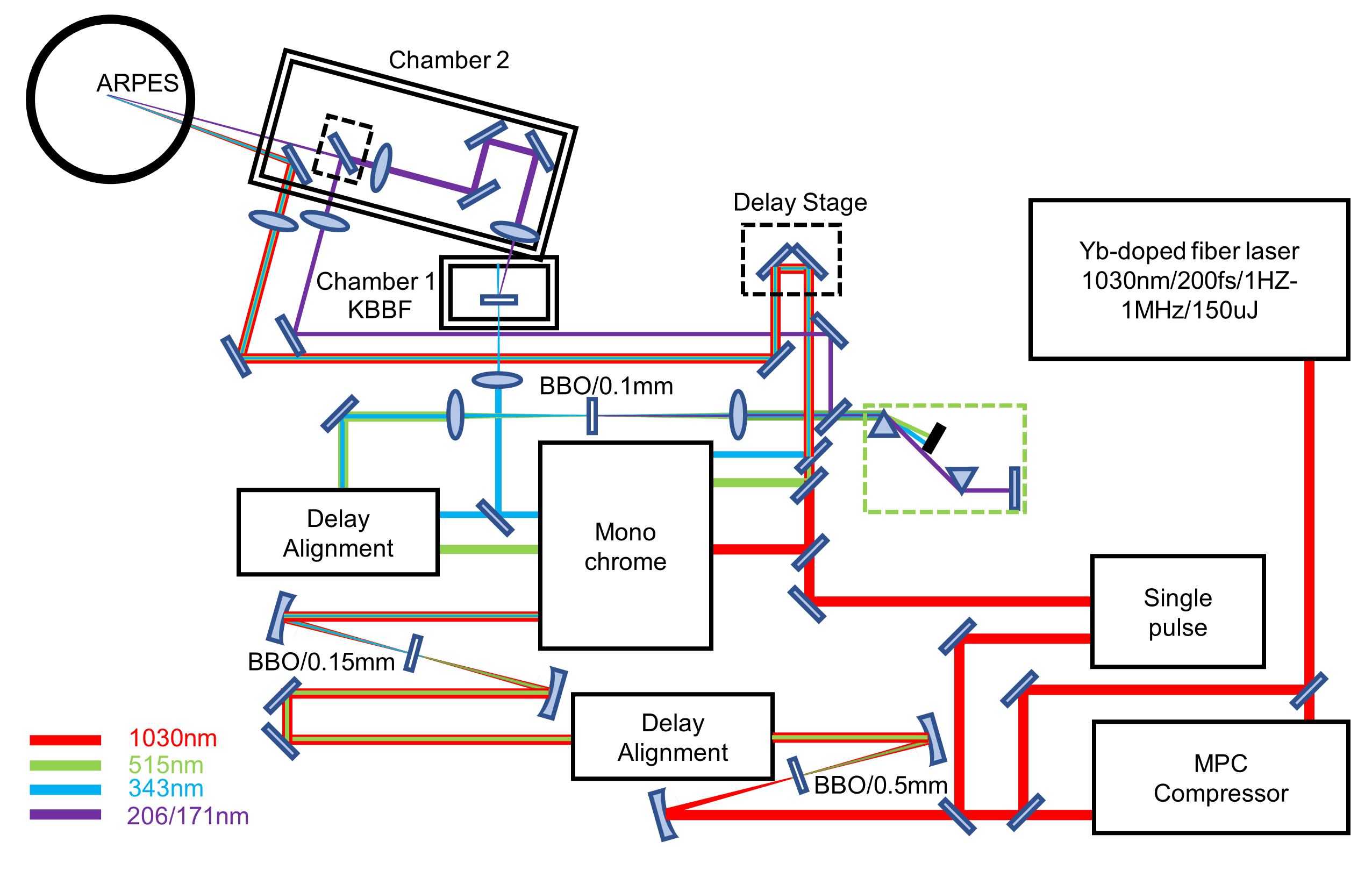}
    \caption{\label{fig:epsart}Schematic diagram of Tr-ARPES with probe photon energy at 6.0 eV and 7.2 eV, as well as pump photon energy at 1.2 eV, 2.4 eV, and 3.6 eV.}
\end{figure*}

\section{Experimental setup}

\subsection{Tr-ARPES layout}
Figure 1 presents a schematic overview of the Tr-ARPES system. The system employs a Yb-doped fiber laser as the fundamental beam (FB) with a central wavelength at 1030 nm ($\rm{h\nu=1.2\ eV}$). The FB provides an adjustable output repetition rate ranging from 1 Hz to 10 MHz, accompanied by a pulse duration of 200 fs and a maximum pulse energy of 150 $\rm{\mu}$J (when operating at or below 300 KHz). 

The FB passes through a setup comprising a half-wave plate and thin film polarizers (TFP), allowing for adjustable splitting power ratios. The transmitted beam subsequently enters the Multi-Pass-Cell (MPC) compressor for secondary compression, resulting in a reduced duration of 75 fs after compression. Subsequently, the compressed beam is combined with the uncompressed beam, which is reflected by the TFP.

Before conducting the experiment, the choice exists between utilizing the compressed beam (CB) with a pulse duration of 75 fs to achieve enhanced time resolution or employing the FB with a pulse duration of 200 fs to enhance energy resolution. Subsequently, the beam is divided using a half-wave plate and TFP before entering the main optical path. This division serves two main purposes: firstly, power reduction, since the total power requirements for the system are not substantial; this reduction helps counteract system instability caused by thermal effects. Secondly, the high-power beam that is reflected can be employed as a dual time-delayed pump with high power density, customized to meet specific experimental requirements.

After entering the main optical path, the beam is focused by a concave mirror onto a BBO crystal (${\theta}$=23°) with a thickness of 0.5 mm, inducing second harmonic (SH) generation at 515 nm  ($\rm{h\nu=2.4\ eV}$). Subsequent to this setup, the residual 1030 nm beam is temporally and spatially aligned with the 515 nm beam precisely. These two beams are then directed onto a BBO crystal (${\theta}$=32.5°) with a thickness of 0.15 mm, resulting in the production of a third harmonic (TH) generation at 343 nm  ($\rm{h\nu=3.6\ eV}$). The efficiency of generating the 343 nm signal can achieve around 30\% under the FB and 18\% under the CB. 

The produced 343 nm, 515 nm, and 1030 nm signals can be effectively monochromatized, yielding a three-way beam consisting of distinct monochromatic components. The 515 nm and 343 nm signals are divided using a beam splitter, while a fraction of the 343 nm signal is directed onto a KBBF crystal to produce a 171 nm ($\rm{h\nu=7.2\ eV}$) signal. The residual 343 nm and 515 nm signals are then focused on a BBO crystal (${\theta}$=77°) with a thickness of 0.1 mm after temporal and spatial alignment, generating a probe beam at 206 nm ($\rm{h\nu=6\ eV}$). 

The surplus portion of the 515 nm and 343 nm signals, along with the additional 1030 nm signal, are combined and directed through a hollow retroreflector (Newport UBBR1-1S) positioned on a delay stage. The merged beam is subsequently focused onto the sample, serving as the pump with variable photon energies. To mitigate pump power, a rotating attenuator is placed in front of the chamber. 

To attain a resolution approaching the Fourier limit (${\Delta}$E * ${\Delta}$T = 1800 fs * meV), the 206 nm signal is subjected to dispersion compensation and is subsequently monochromatized using a pair of prisms before entering the chamber. Conversely, for the probe beam of 171 nm, a pre-compensation scheme is utilized to alleviate the dispersion effects induced by the 343 nm signal, owing to the challenges associated with direct dispersion compensation.

Our ARPES system is equipped with the following components: (1) Hemispherical analyzer (Scienta DA30L), (2) Helium lamp system (Scienta VUV5050) with photon energies of 21.2 eV and 40.8 eV, offering an energy resolution of less than 5 meV, (3) 6-axis sample manipulator (Fermion instrument) capable of operating at temperature as low as 6 K, (4) Annealing chamber featuring an argon gun (Speces), radiation capability (up to 700 ${^\circ}$C), and laser heating capacity (up to 1200 ${^\circ}$C), enabling fundamental sample processing, and (5) Our internally developed Delay Stage software, seamlessly integrate with the analyzer software SES, facilitating automated spectrum sweeping at different delay times. Moreover, our main ARPES chamber achieves an exceptional vacuum level,  surpassing even $3\times10^{-11}$ Torr following meticulous baking procedures.

\subsection{MPC pulse compress}
The pulse duration of a UV laser is commonly influenced by the pulse duration of the incident beam. Employing a shorter beam pulse duration enhances time resolution in Tr-ARPES. Additionally, the reduced pulse duration enables higher peak power density for the pump, thereby facilitating processes such as optical modulation and photo-induced phase transitions. As a result, we employ secondary compression techniques to achieve a shorter pulse duration.

When there is no dispersion present, the product of the pulse's spectral width and duration remains constant. This implies that shorter laser pulses exhibit broader spectral widths. Thus, in order to achieve shorter laser pulses, it is necessary to broaden the pulse spectrum to encompass the desired pulse duration. Nonlinear self-phase modulation within a nonlinear medium is a commonly employed method for achieving spectral broadening. Among the various techniques for pulse width compression, hollow fiber compression and Multi-Pass Cell (MPC) compression \cite{lavenu2018nonlinear,weitenberg2017nonlinear} have emerged as the most widely adopted strategies globally. Comparatively, the MPC scheme offers several advantages over the hollow fiber compression scheme, including high efficiency, a compact footprint, and enhanced stability. Consequently, we have chosen the MPC compression scheme and have successfully reduced the pulse duration from 200 fs to 75 fs, achieving an efficiency that exceeds 75\%.

Figure 2a illustrates the optical path of the MPC. A 10W FB laser with a duration of 200fs at a repetition rate of 500 kHz is directed into the MPC through a D-shaped reflection mirror. The MPC system is comprised of two 2-inch concave mirrors, each with a focal length of 250 mm. Positioned between these mirrors is a fused silica with a thickness of 3 mm and an element coated with a double-sided permeability-enhancing film for 1030nm. 

During its course through the MPC, the beam undergoes 36 reflections, with 18 beam drop-off points on the mirrors arranged in a circular pattern. Following these reflections, the spectrally broadened beam exits through another D-shaped reflection mirror. Passing through the fused silica element in each round introduces 100 fs$^2$ of group delay dispersion (GDD), resulting in an accumulated dispersion of approximately 3600 fs$^2$. 

Furthermore, the MPC itself contributes a certain level of positive dispersion. To counteract this and the dispersion arising from the broadened spectrum, chirped mirrors are employed. These mirrors have a combined dispersion value of -5000 fs$^2$. The compression capabilities of an individual MPC exhibit limitations. To attain further pulse duration compression, an additional round of MPC compression can be executed. Following dispersion compensation, we achieve a CB power of 8 W with an efficiency of 80\%. Figure 2(b-d) presents the spectral composition before and after compression, along with the ensuing alteration in pulse duration.

\begin{figure}[ht]
    \centering
    \includegraphics[width=0.48\textwidth]{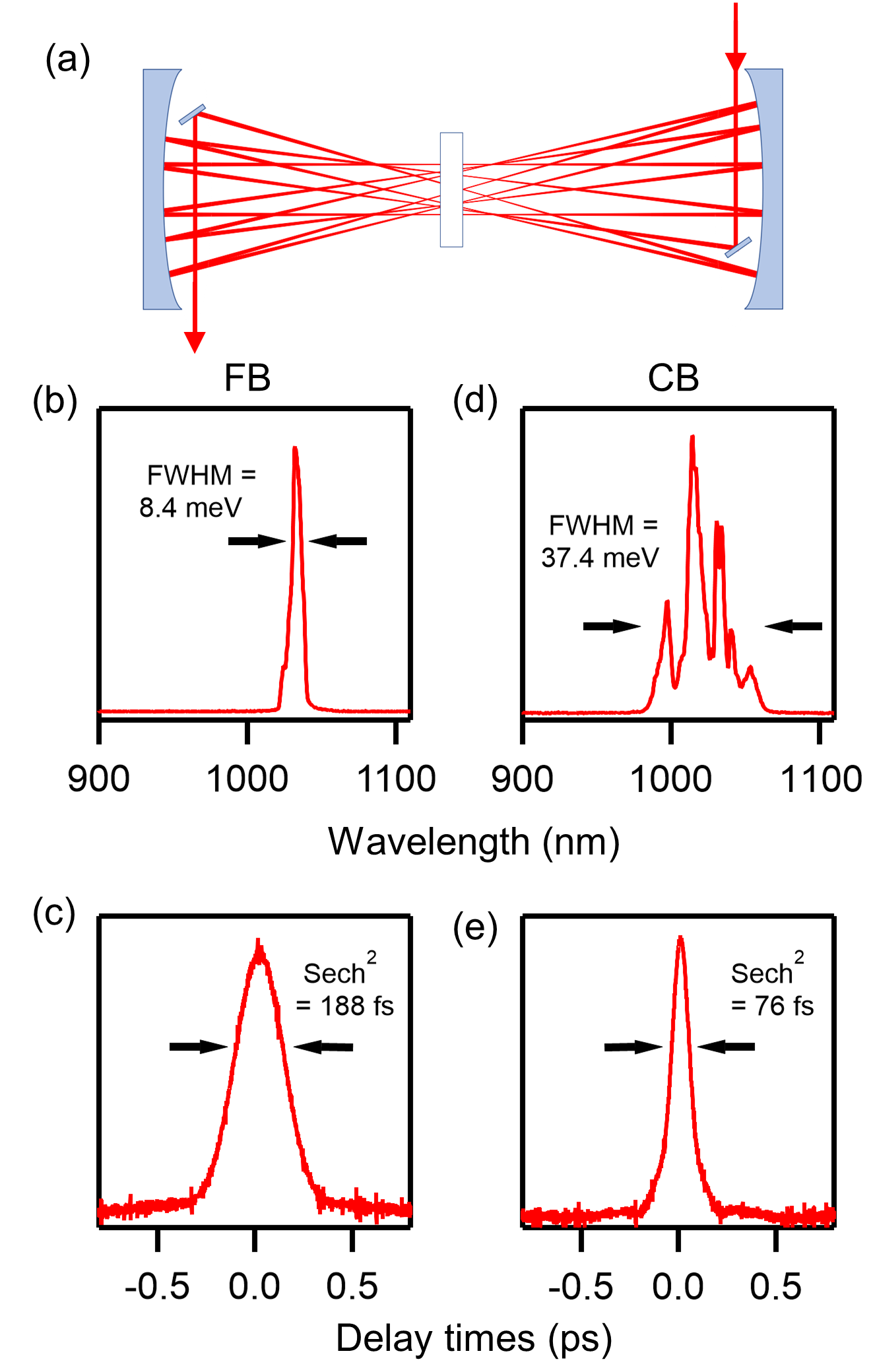}
    \caption{\label{fig:epsart} MPC Diagram and Spectra/Duration Results (a): MPC Diagram. (b-c): Spectra and Duration Before MPC. (d-e): Spectra and Duration After MPC. }
\end{figure}

\begin{figure*}[ht]
    \centering
    \includegraphics[width=1\textwidth]{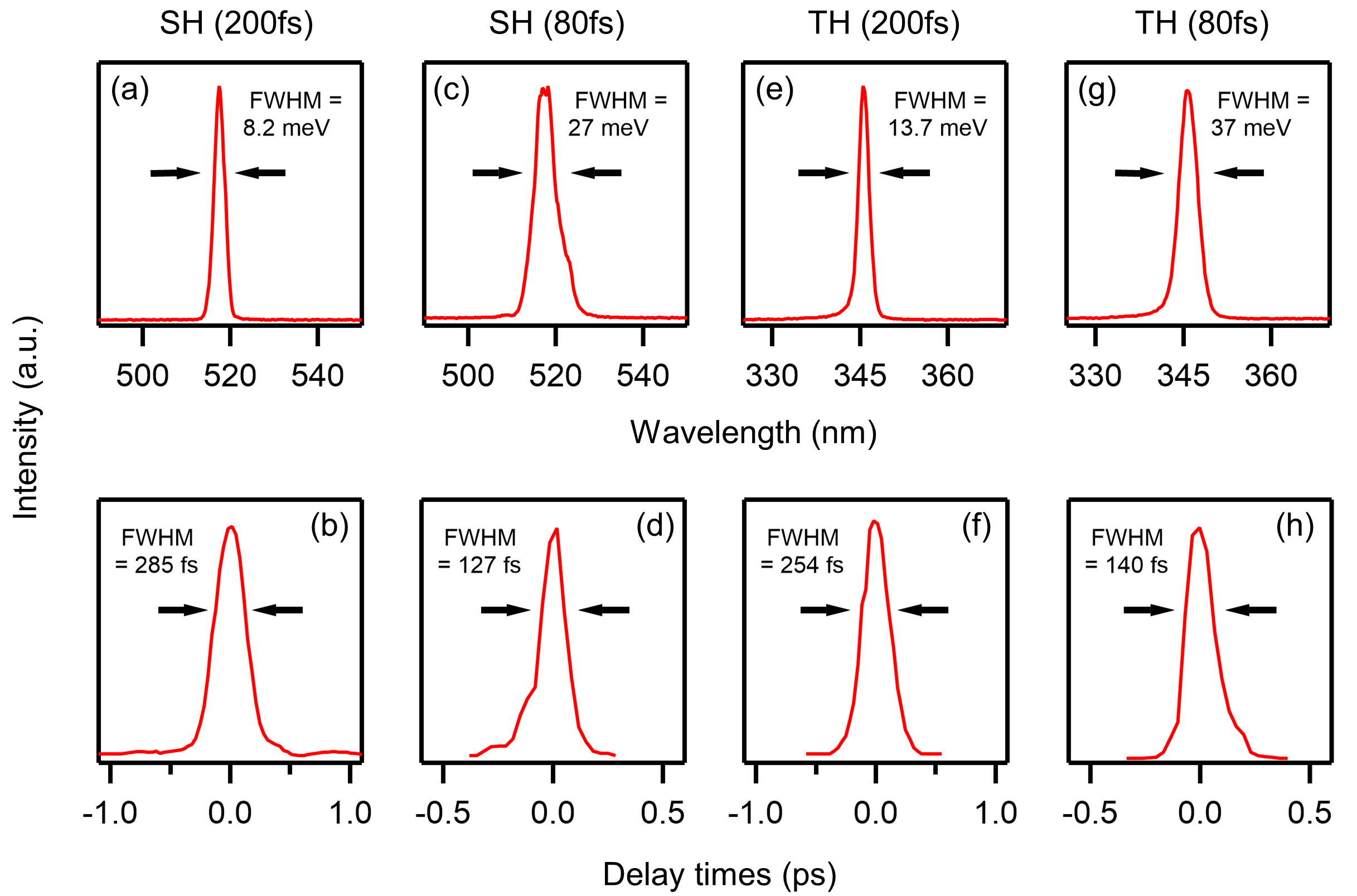}
    \caption{\label{fig:epsart} Spectra and cross-correlation data for 515 nm (a-d) and 343 nm (e-h) before and after MPC pulse compression. }
\end{figure*}

\subsection{Character of pump and probe}
Figure 3 illustrates the spectra and cross-correlation results of 515 nm and 343 nm before (a-b), (e-f), and after (c-d), (g-h) compression.

To calibrate the spot size of the pump pulses on the analyzer focus, we measure the pump power behind the sample position by scanning the spot along the sharp edge of the gold (Au) sample. The Gaussian curve can be derived through the differentiation of this power curve, with the full width at half maximum (FWHM) of the Gaussian curve representing the size of the pump laser. Since measuring the probe intensity behind the sample is unfeasible, we choose to use the curve that illustrates the intensity changes of the counts on the analyzer as the sample position is adjusted. This measurement was carried out with the sample surface placed at a 45-degree angle to the incident laser. Figure 4 illustrates a spot size of 6 eV, which is approximately 46 $\rm{\mu}$m $\times$ 25 $\rm{\mu}$m. After compensating for the lateral spread resulting from the 45-degree incidence angle, the spot size is approximately 32 $\rm{\mu}$m $\times$ 25 $\rm{\mu}$m. Further details regarding the spot size of the remaining pump and probe can be found in Table I.

\begin{figure}[ht]
    \centering
    \includegraphics[width=0.5\textwidth]{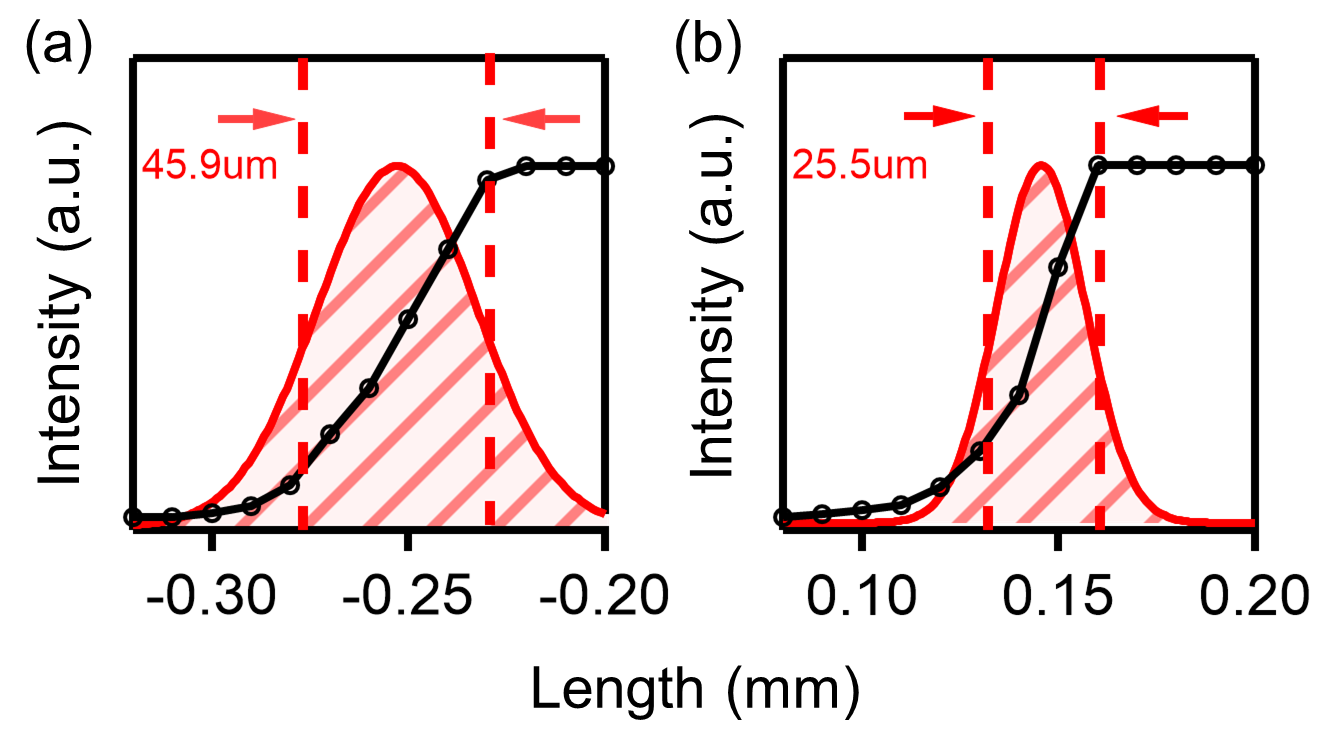}
    \caption{\label{fig:epsart} Spot size at 6 eV with a 45-degree incidence angle. (a) Horizontal Spot Size. (b) Vertical Spot Size.}
\end{figure}

\begin{table}[ht]
    \caption{\ Spot size of pump and probe }
    \label{table:SpotSize}
    \begin{tabular}{p{3cm}p{3cm}p{1.7cm}}
        \hline
        \hline
        h$\nu$(eV) & Horizon(um) & Vertical(um)\\
        \hline
        $\rm{1.2\ eV}$ & 107.2 & 64.3 \\
        $\rm{2.4\ eV}$ & 75.5 & 61.3 \\
        $\rm{3.6\ eV}$ & 45.4 & 88 \\
        $\rm{6\ eV}$ & 45.9 & 25.5 \\
        $\rm{7.2\ eV}$ & 23.0 & 22.56 \\
        \hline
    \end{tabular}
\end{table}

\begin{figure*}
    \centering
    \includegraphics[width=1\textwidth]{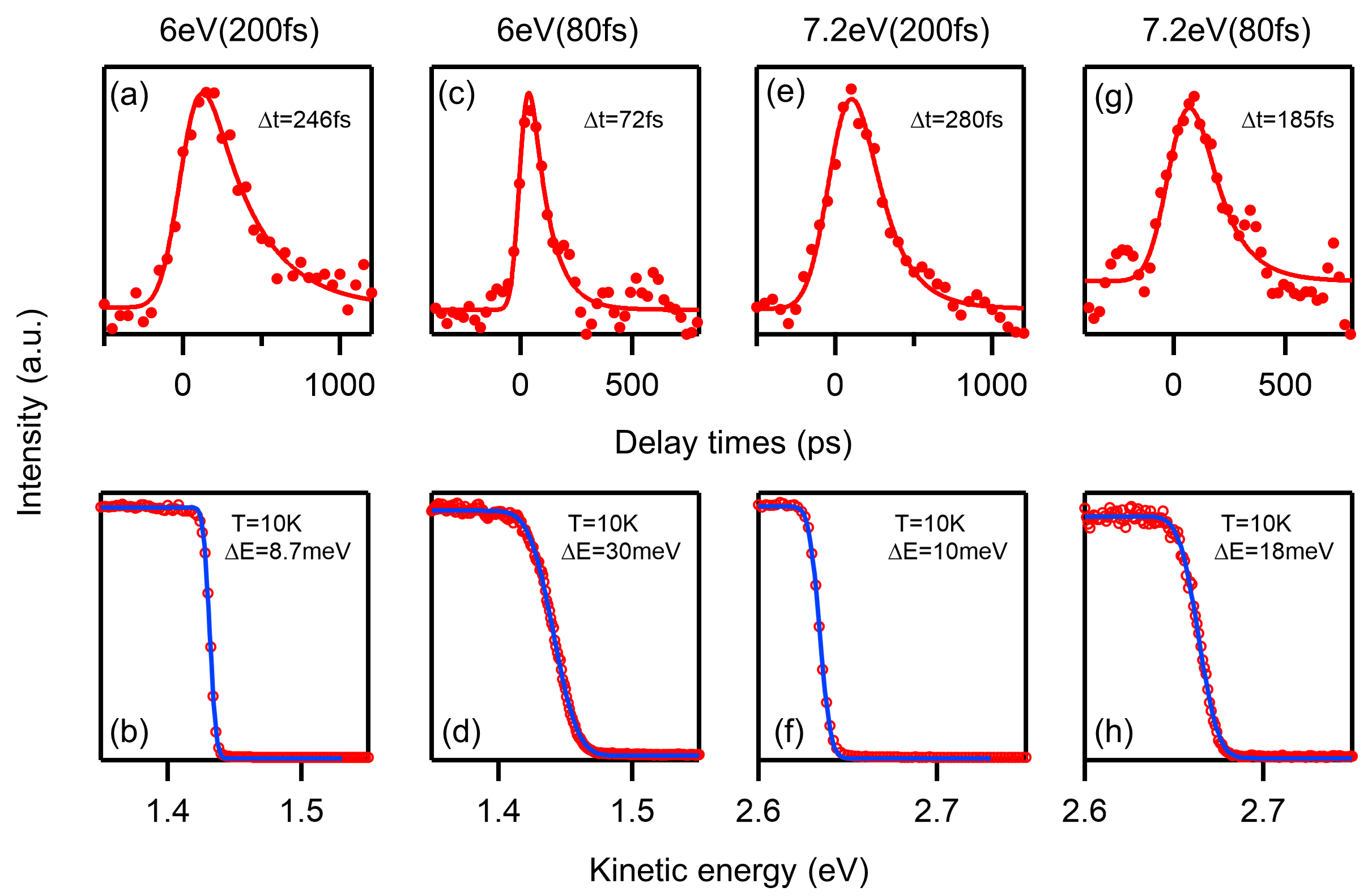}
    \caption{\label{fig:epsart} Energy and Time Resolution Under Various Configurations. Energy resolution is determined for polycrystalline gold at a temperature of 10 K, while time resolution is evaluated using a 2.4 eV pump. (a-d): Raw and fitted data illustrating time and energy resolution of 6 eV at 200 fs and 80 fs. (e-h): Raw and fitted data demonstrating time and energy resolution of 7.2 eV at 200 fs and 80 fs.}
\end{figure*}

\section{Energy resolution and time resolution}
The energy resolution is obtained through the measurement of the Fermi edge of polycrystalline gold at 10K. This measurement entails fitting a convolution of a Gaussian distribution ${g\left( \omega  \right) = {e^{\left( { - 4\ln 2} \right){\omega ^2}/R_e^2}}}$ and the Fermi-Dirac distribution ${f\left( \omega  \right) = 1/\left( {{e^{\hbar \omega /{k_B}T}} + 1} \right)}$ to the data, where the FWHM of the Gaussian distribution characterizes the energy resolution. Time resolution measurement involves fitting the temporal distribution curve of the density of states in the high-energy region of an HOPG sample. The fitting function is the product of the Heaviside function and a single-exponential function convolved with a Gaussian function \cite{bao2021full}.
\[I\left( t \right) = A\left( {1 + \mathrm{erf}\left( {\frac{{t - {t_0}}}{{\Delta t}}} \right)} \right){e^{ - \frac{{t - {t_o}}}{\tau }}} + B\]
Here, the parameter ${\Delta t}$ corresponds to the time resolution, while ${\tau }$ corresponds to the relaxation time of decay.

Figure 5 illustrates the energy and time resolution curves for the 2.4 eV pump under both 80 fs and 200 fs beams. Table II presents the values for time and energy resolution across varying pump photon energies. During excitation with a 1.2 eV pump, the time resolution is slightly worse than that with a 2.4 eV pump. This difference can be attributed to the broader pulse duration of the 1.2 eV pump in comparison to the 2.4 eV pump.

\begin{table}[ht]
    \caption{\ Different Resolutions Under Various Configurations }
    \label{table:Resolution}
    \begin{tabular}{p{2.2cm}p{2.2cm}p{2.2cm}p{1cm}}
        \hline
        \hline
        \  & $\Delta t$\ 1.2\ eV & $\Delta t$\ 2.4\ eV & $\Delta E$\\
        \hline
        $\rm{6\ eV(80fs)}$ & 95fs & 72fs & 30meV \\
        $\rm{7.2\ eV(80fs)}$ & 195fs & 185fs & 18meV \\
        $\rm{6\ eV(200fs)}$ & 290fs & 246fs & 8.7meV \\
        $\rm{7.2\ eV(200fs)}$ & 300fs & 280fs & 10meV \\
        \hline
    \end{tabular}
\end{table}

\section{Measurement}
\subsection{Equilibrium state of  $\mathbf{Bi_2Se_3}$}

Figure 6 depicts the measurement of the equilibrium state of $\rm{Bi_2Se_3}$ at the $\Gamma$-point using photon energies of 6 eV and 7.2 eV at the temperature of 10K. Owing to the higher photon energy of 7.2 eV, a larger momentum space can be observed within the same acceptant angle range. According to the formula ${k_\parallel } = \sqrt {2m{E_{kin}}} \rm{sin}\theta /\hbar $, the maximum observable momentum k at 7.2 eV is 0.84$(1/\text {Å})$, with a momentum k of 0.22$(1/\text {Å})$ at a 15-degree angle. For 6 eV, the maximum observable momentum k is 0.62$(1/\text {Å})$, with a momentum k of 0.16$(1/\text {Å})$ at a 15-degree angle. With an equal number of analyzer channels, 6 eV provides improved angular resolution. Furthermore, since measurements at 6 eV and 7.2 eV capture distinct kz values, it becomes feasible to investigate band dispersion at diverse kz values in specific materials \cite{zhong2023nodeless}.

\begin{figure}[ht]
    \centering
    \includegraphics[width=0.48\textwidth]{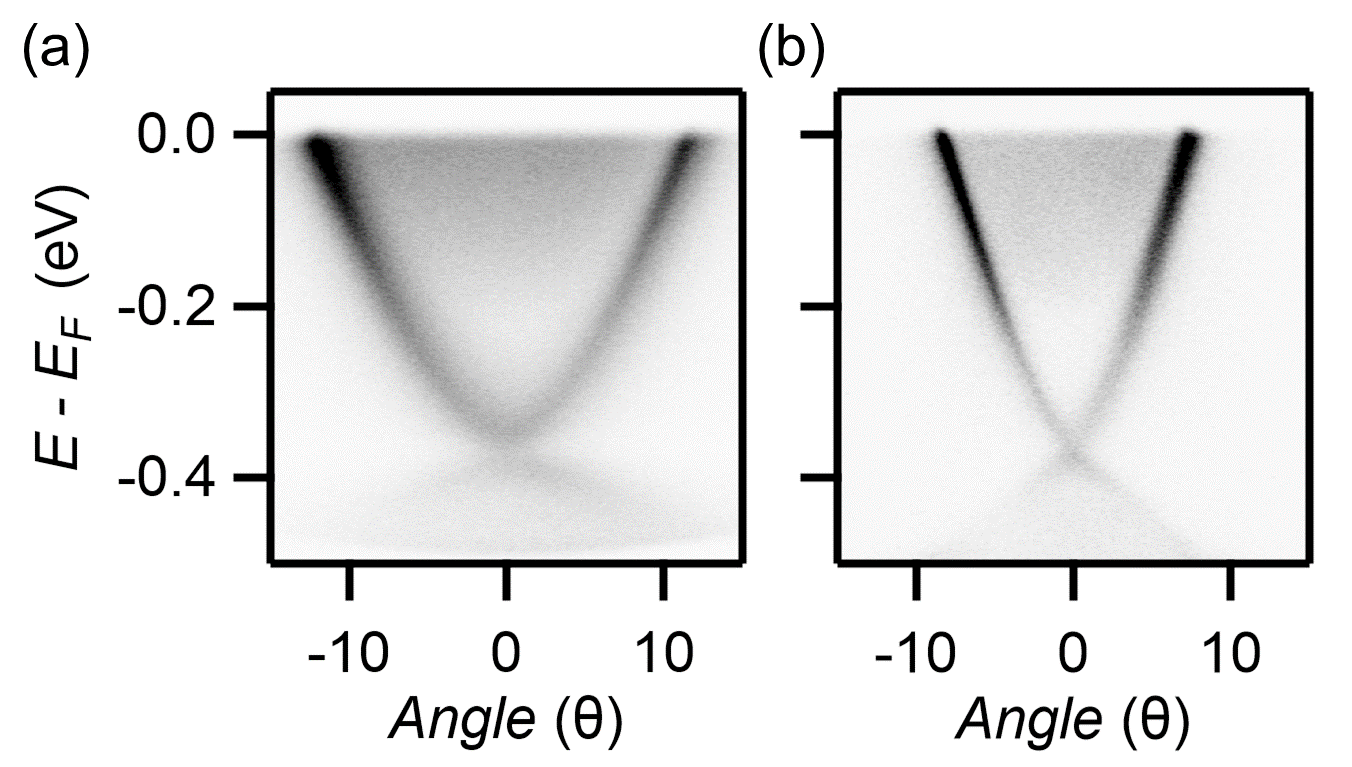}
    \caption{\label{fig:epsart} Depiction of Equilibrium State of Bi2Se3 in Tr-ARPES. (a) Band structure of $\rm{Bi_2Se_3}$ at the $\Gamma$ point measured by 6 eV. (b) Band structure of $\rm{Bi_2Se_3}$ at the $\Gamma$ point measured by 7.2 eV.}
\end{figure}

\subsection{Measurement of the gap in  $\mathbf{MnBi_2Te_4}$.}
To validate the energy resolution capability of our equipment, we chose the topological material $\rm{MnBi_2Te_4}$. Previous research has indicated the existence of a gap in the topological surface state of $\rm{MnBi_2Te_4}$ material, with an approximate gap size of 10 meV \cite{li2019dirac}. We conducted band structure measurements at the $\Gamma$-point of $\rm{MnBi_2Te_4}$ by employing the 7.2 eV probe laser with configuration of optimized energy resolution at 10meV. Figure 7a illustrates the band structure results at the $\Gamma$-point, while Figure 7b presents the second derivative of the data. We performed an analysis of individual Energy Distribution Curves (EDCs) at various momentum k values, as shown in Figure 7c. The arrows in the figure indicate the peak positions of the EDCs at different momenta, corresponding to the gap size. Through the examination of these peak positions, we confirmed that the gap at the $\Gamma$-point is around 10 meV, thus validating the energy resolution capability of our equipment.

\begin{figure}[ht]
    \centering
    \includegraphics[width=0.48\textwidth]{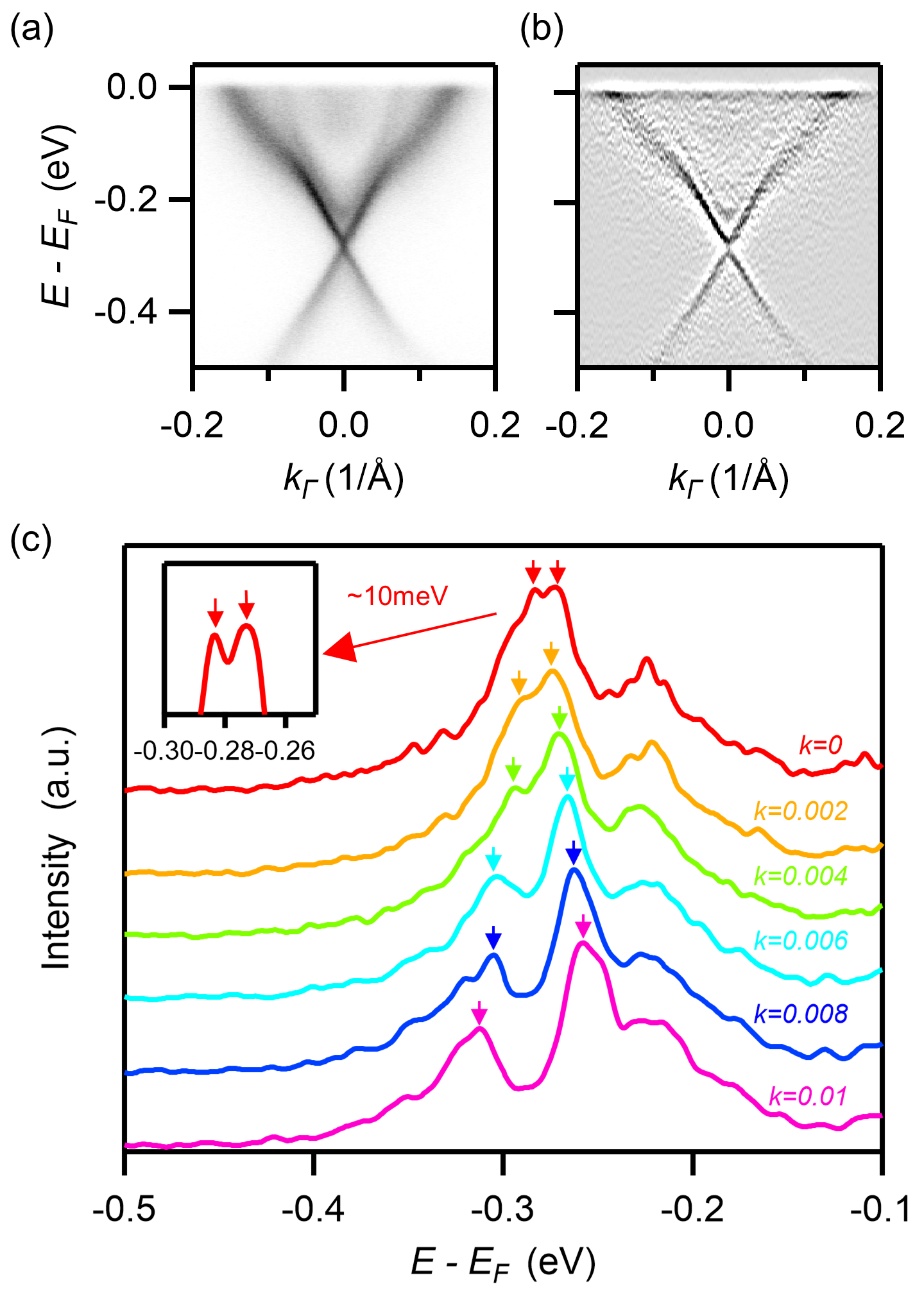}
    \caption{\label{fig:epsart} Measurement Results of $\rm{MnBi_2Te_4}$ with the configuration of better energy resolution. (a) Band structure at the $\Gamma$ point.(b) Second derivative of (a). (c) Energy distribution curves (EDCs) at different k points from (a), with peak positions indicated by arrows.}
\end{figure}

\subsection{Dynamic of $\mathbf{Ta_2NiSe_5}$}
We selected the excitonic insulator candidate $\rm{Ta_2NiSe_5}$ to validate the temporal resolution capabilities of our experimental setup. In this material, a hole-like pocket exists at the center of the Brillouin zone and a gap opens below the transition temperature \cite{Wakisaka2009, Mor2017, Fukutani2021}. In Figure 8a, we present the static band structure of the hole band at the $\Gamma$-point measured by a 6eV probe laser with an optimized time resolution of 95fs. Upon excitation with 1.2 eV pump laser at fluence of $0.25 \mathrm{~mJ} / \mathrm{cm}^2$, Figure 8b illustrates the photoemission intensity as a function of energy and time delay along the red dashed line in panel (a). The temporal evolution of the integrated intensity is shown in the lower part of Figure 8c, corresponding to the region outlined by the white dashed line in panel (b). The fitted blue solid curve represents the incoherent component within the data, modeled as the product of the Heaviside function and a single-exponential function convolved with a Gaussian function. After subtracting the incoherent contribution, the upper part of Figure 8c shows a clear oscillation. Through Fourier transformation, we determined a frequency of approximately 1.0 THz, consistent with the $A_g$ phonon mode \cite{Werdehausen2018, Tang2020}. This result highlights the capability of our apparatus to capture fast dynamic phenomena within materials, encompassing collective mode oscillations and rapid interaction processes.

\begin{figure}[ht]
    \centering
    \includegraphics[width=0.48\textwidth]{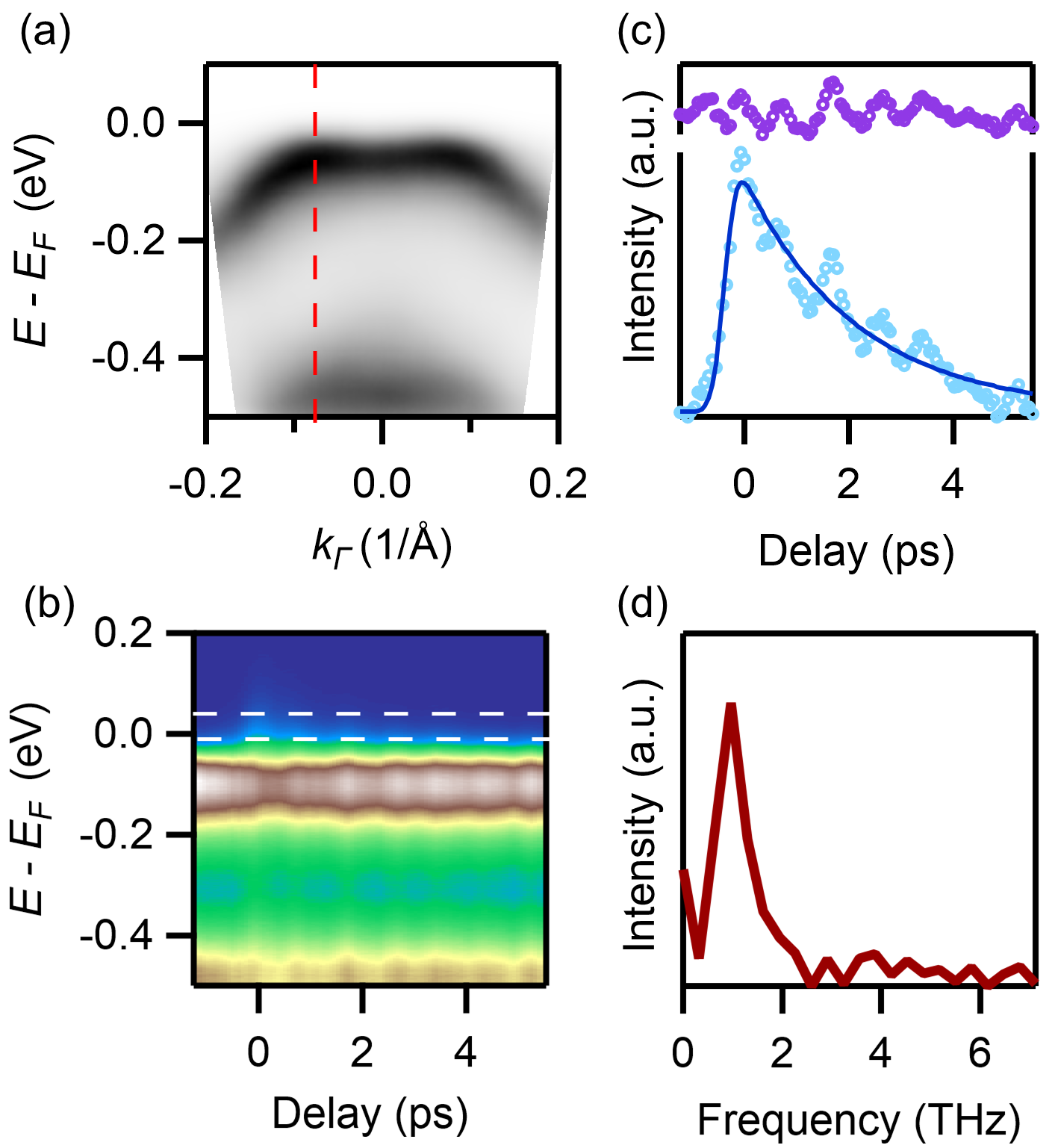}
    \caption{\label{fig:epsart} Measurement Results of $\rm{Ta_2NiSe_5}$. (a) Band structure at the $\Gamma$ point.(b) Photoemission intensity as a function of energy and time delay at the red dashed line cut in panel (a). (c) Bottom: The integrated intensity as a function of time delay at the region of the white dashed line in panel (b). The blue curve indicates the incoherent part fitted by the product of the Heaviside function and a single-exponential function convolved with a Gaussian function. Top: Intensity after removing the incoherent part. (d) The corresponding Fourier transform magnitude of the curve in the top of panel (c).}
\end{figure}

\section{Conclusion}
In summary, we have successfully developed a time-resolved ARPES system employing a 200 fs Yb-doped laser coupled with a hemispherical energy analyzer. The photon energy of the pump laser can be selectively tuned to 1.2 eV, 2.4 eV, and 3.6 eV, while the photon energy of the probe laser is adjustable to 6 eV and 7.2 eV. Through the application of MPC compression to the laser, we have achieved 75 fs CB. The use of the 200 fs FB configuration has notably improved the energy resolution, and alternatively, employing the 80 fs CB configuration has provided remarkable enhancements in time resolution. The enhanced energy resolution allows us to discern intricate features within specific band structures, as evidenced by our successful verification of a gap size of approximately 10 meV in the surface state of the topological material $\rm{MnBi_2Te_4}$. Moreover, the improved time resolution empowers us to explore material dynamics, as evidenced by our successful characterization of oscillations in the $\rm{Ta_2NiSe_5}$. Our innovative system presents an adaptable platform for future experiments, readily customizable to meet diverse research requirements.

\begin{acknowledgments}
We thank H.C. Lei and S.J. Tian from Renmin University(Beijing) for providing high quality $\rm{MnBi_2Te_4}$ samples, Z.W. Wang from Beijing Institute of Technology for providing $\rm{Ta_2NiSe_5}$ samples, S.Y. Xu from Northwest University(Xi’an, Shaanxi) for providing optical guidance.

This work was supported by the Ministry of Science and Technology of China (Grant No. 2022YFA1403800),  the National Natural Science Foundation of China (Grant Nos. U1832202 and 11888101), the Beijing Municipal Science and Technology Commission (Grant No. Z171100002017018), the Chinese Academy of Sciences (Grant Nos. XDB33000000 and XDB28000000), the Informatization Plan of Chinese Academy of Sciences (Grant No.CAS-WX2021SF-0102), and the Synergetic Extreme Condition User Facility (SECUF).

Mojun Pan, Junde Liu and Famin Chen contributed equally to this work.

\end{acknowledgments}

\bibliography{aipsamp}% Produces the bibliography via BibTeX.

\providecommand{\noopsort}[1]{}\providecommand{\singleletter}[1]{#1}%
\begin{thebibliography}{40}%
\makeatletter
\providecommand \@ifxundefined [1]{%
 \@ifx{#1\undefined}
}%
\providecommand \@ifnum [1]{%
 \ifnum #1\expandafter \@firstoftwo
 \else \expandafter \@secondoftwo
 \fi
}%
\providecommand \@ifx [1]{%
 \ifx #1\expandafter \@firstoftwo
 \else \expandafter \@secondoftwo
 \fi
}%
\providecommand \natexlab [1]{#1}%
\providecommand \enquote  [1]{``#1''}%
\providecommand \bibnamefont  [1]{#1}%
\providecommand \bibfnamefont [1]{#1}%
\providecommand \citenamefont [1]{#1}%
\providecommand \href@noop [0]{\@secondoftwo}%
\providecommand \href [0]{\begingroup \@sanitize@url \@href}%
\providecommand \@href[1]{\@@startlink{#1}\@@href}%
\providecommand \@@href[1]{\endgroup#1\@@endlink}%
\providecommand \@sanitize@url [0]{\catcode `\\12\catcode `\$12\catcode
  `\&12\catcode `\#12\catcode `\^12\catcode `\_12\catcode `\%12\relax}%
\providecommand \@@startlink[1]{}%
\providecommand \@@endlink[0]{}%
\providecommand \url  [0]{\begingroup\@sanitize@url \@url }%
\providecommand \@url [1]{\endgroup\@href {#1}{\urlprefix }}%
\providecommand \urlprefix  [0]{URL }%
\providecommand \Eprint [0]{\href }%
\providecommand \doibase [0]{http://dx.doi.org/}%
\providecommand \selectlanguage [0]{\@gobble}%
\providecommand \bibinfo  [0]{\@secondoftwo}%
\providecommand \bibfield  [0]{\@secondoftwo}%
\providecommand \translation [1]{[#1]}%
\providecommand \BibitemOpen [0]{}%
\providecommand \bibitemStop [0]{}%
\providecommand \bibitemNoStop [0]{.\EOS\space}%
\providecommand \EOS [0]{\spacefactor3000\relax}%
\providecommand \BibitemShut  [1]{\csname bibitem#1\endcsname}%
\let\auto@bib@innerbib\@empty
%</preamble>
\bibitem [{\citenamefont {Wang}\ \emph {et~al.}(2012)\citenamefont {Wang},
  \citenamefont {Hsieh}, \citenamefont {Sie}, \citenamefont {Steinberg},
  \citenamefont {Gardner}, \citenamefont {Lee}, \citenamefont
  {Jarillo-Herrero},\ and\ \citenamefont {Gedik}}]{wang2012measurement}%
  \BibitemOpen
  \bibfield  {author} {\bibinfo {author} {\bibfnamefont {Y.}~\bibnamefont
  {Wang}}, \bibinfo {author} {\bibfnamefont {D.}~\bibnamefont {Hsieh}},
  \bibinfo {author} {\bibfnamefont {E.}~\bibnamefont {Sie}}, \bibinfo {author}
  {\bibfnamefont {H.}~\bibnamefont {Steinberg}}, \bibinfo {author}
  {\bibfnamefont {D.}~\bibnamefont {Gardner}}, \bibinfo {author} {\bibfnamefont
  {Y.}~\bibnamefont {Lee}}, \bibinfo {author} {\bibfnamefont {P.}~\bibnamefont
  {Jarillo-Herrero}}, \ and\ \bibinfo {author} {\bibfnamefont {N.}~\bibnamefont
  {Gedik}},\ }\href@noop {} {\bibfield  {journal} {\bibinfo  {journal} {Phys.
  Rev. Lett.}\ }\textbf {\bibinfo {volume} {109}},\ \bibinfo {pages} {127401}
  (\bibinfo {year} {2012})}\BibitemShut {NoStop}%
\bibitem [{\citenamefont {Sobota}\ \emph {et~al.}(2012)\citenamefont {Sobota},
  \citenamefont {Yang}, \citenamefont {Analytis}, \citenamefont {Chen},
  \citenamefont {Fisher}, \citenamefont {Kirchmann},\ and\ \citenamefont
  {Shen}}]{sobota2012ultrafast}%
  \BibitemOpen
  \bibfield  {author} {\bibinfo {author} {\bibfnamefont {J.~A.}\ \bibnamefont
  {Sobota}}, \bibinfo {author} {\bibfnamefont {S.}~\bibnamefont {Yang}},
  \bibinfo {author} {\bibfnamefont {J.~G.}\ \bibnamefont {Analytis}}, \bibinfo
  {author} {\bibfnamefont {Y.}~\bibnamefont {Chen}}, \bibinfo {author}
  {\bibfnamefont {I.~R.}\ \bibnamefont {Fisher}}, \bibinfo {author}
  {\bibfnamefont {P.~S.}\ \bibnamefont {Kirchmann}}, \ and\ \bibinfo {author}
  {\bibfnamefont {Z.-X.}\ \bibnamefont {Shen}},\ }\href@noop {} {\bibfield
  {journal} {\bibinfo  {journal} {Phys. Rev. Lett.}\ }\textbf {\bibinfo
  {volume} {108}},\ \bibinfo {pages} {117403} (\bibinfo {year}
  {2012})}\BibitemShut {NoStop}%
\bibitem [{\citenamefont {S{\'a}nchez-Barriga}\ \emph
  {et~al.}(2016)\citenamefont {S{\'a}nchez-Barriga}, \citenamefont {Golias},
  \citenamefont {Varykhalov}, \citenamefont {Braun}, \citenamefont {Yashina},
  \citenamefont {Schumann}, \citenamefont {Min{\'a}r}, \citenamefont {Ebert},
  \citenamefont {Kornilov},\ and\ \citenamefont
  {Rader}}]{sanchez2016ultrafast}%
  \BibitemOpen
  \bibfield  {author} {\bibinfo {author} {\bibfnamefont {J.}~\bibnamefont
  {S{\'a}nchez-Barriga}}, \bibinfo {author} {\bibfnamefont {E.}~\bibnamefont
  {Golias}}, \bibinfo {author} {\bibfnamefont {A.}~\bibnamefont {Varykhalov}},
  \bibinfo {author} {\bibfnamefont {J.}~\bibnamefont {Braun}}, \bibinfo
  {author} {\bibfnamefont {L.}~\bibnamefont {Yashina}}, \bibinfo {author}
  {\bibfnamefont {R.}~\bibnamefont {Schumann}}, \bibinfo {author}
  {\bibfnamefont {J.}~\bibnamefont {Min{\'a}r}}, \bibinfo {author}
  {\bibfnamefont {H.}~\bibnamefont {Ebert}}, \bibinfo {author} {\bibfnamefont
  {O.}~\bibnamefont {Kornilov}}, \ and\ \bibinfo {author} {\bibfnamefont
  {O.}~\bibnamefont {Rader}},\ }\href@noop {} {\bibfield  {journal} {\bibinfo
  {journal} {Phys. Rev. B}\ }\textbf {\bibinfo {volume} {93}},\ \bibinfo
  {pages} {155426} (\bibinfo {year} {2016})}\BibitemShut {NoStop}%
\bibitem [{\citenamefont {Ciocys}\ \emph {et~al.}(2020)\citenamefont {Ciocys},
  \citenamefont {Morimoto}, \citenamefont {Mori}, \citenamefont {Gotlieb},
  \citenamefont {Hussain}, \citenamefont {Analytis}, \citenamefont {Moore},\
  and\ \citenamefont {Lanzara}}]{ciocys2020manipulating}%
  \BibitemOpen
  \bibfield  {author} {\bibinfo {author} {\bibfnamefont {S.}~\bibnamefont
  {Ciocys}}, \bibinfo {author} {\bibfnamefont {T.}~\bibnamefont {Morimoto}},
  \bibinfo {author} {\bibfnamefont {R.}~\bibnamefont {Mori}}, \bibinfo {author}
  {\bibfnamefont {K.}~\bibnamefont {Gotlieb}}, \bibinfo {author} {\bibfnamefont
  {Z.}~\bibnamefont {Hussain}}, \bibinfo {author} {\bibfnamefont {J.~G.}\
  \bibnamefont {Analytis}}, \bibinfo {author} {\bibfnamefont {J.~E.}\
  \bibnamefont {Moore}}, \ and\ \bibinfo {author} {\bibfnamefont
  {A.}~\bibnamefont {Lanzara}},\ }\href@noop {} {\bibfield  {journal} {\bibinfo
   {journal} {npj Quantum Materials}\ }\textbf {\bibinfo {volume} {5}},\
  \bibinfo {pages} {16} (\bibinfo {year} {2020})}\BibitemShut {NoStop}%
\bibitem [{\citenamefont {Sobota}\ \emph {et~al.}(2014)\citenamefont {Sobota},
  \citenamefont {Yang}, \citenamefont {Leuenberger}, \citenamefont {Kemper},
  \citenamefont {Analytis}, \citenamefont {Fisher}, \citenamefont {Kirchmann},
  \citenamefont {Devereaux},\ and\ \citenamefont {Shen}}]{sobota2014ultrafast}%
  \BibitemOpen
  \bibfield  {author} {\bibinfo {author} {\bibfnamefont {J.~A.}\ \bibnamefont
  {Sobota}}, \bibinfo {author} {\bibfnamefont {S.-L.}\ \bibnamefont {Yang}},
  \bibinfo {author} {\bibfnamefont {D.}~\bibnamefont {Leuenberger}}, \bibinfo
  {author} {\bibfnamefont {A.~F.}\ \bibnamefont {Kemper}}, \bibinfo {author}
  {\bibfnamefont {J.~G.}\ \bibnamefont {Analytis}}, \bibinfo {author}
  {\bibfnamefont {I.~R.}\ \bibnamefont {Fisher}}, \bibinfo {author}
  {\bibfnamefont {P.~S.}\ \bibnamefont {Kirchmann}}, \bibinfo {author}
  {\bibfnamefont {T.~P.}\ \bibnamefont {Devereaux}}, \ and\ \bibinfo {author}
  {\bibfnamefont {Z.-X.}\ \bibnamefont {Shen}},\ }\href@noop {} {\bibfield
  {journal} {\bibinfo  {journal} {J. Electron Spectrosc. Relat. Phenom.}\
  }\textbf {\bibinfo {volume} {195}},\ \bibinfo {pages} {249} (\bibinfo {year}
  {2014})}\BibitemShut {NoStop}%
\bibitem [{\citenamefont {Damascelli}, \citenamefont {Hussain},\ and\
  \citenamefont {Shen}(2003)}]{damascelli2003angle}%
  \BibitemOpen
  \bibfield  {author} {\bibinfo {author} {\bibfnamefont {A.}~\bibnamefont
  {Damascelli}}, \bibinfo {author} {\bibfnamefont {Z.}~\bibnamefont {Hussain}},
  \ and\ \bibinfo {author} {\bibfnamefont {Z.-X.}\ \bibnamefont {Shen}},\
  }\href@noop {} {\bibfield  {journal} {\bibinfo  {journal} {Rev. Mod. Phys.}\
  }\textbf {\bibinfo {volume} {75}},\ \bibinfo {pages} {473} (\bibinfo {year}
  {2003})}\BibitemShut {NoStop}%
\bibitem [{\citenamefont {Perfetti}\ \emph {et~al.}(2007)\citenamefont
  {Perfetti}, \citenamefont {Loukakos}, \citenamefont {Lisowski}, \citenamefont
  {Bovensiepen}, \citenamefont {Eisaki},\ and\ \citenamefont
  {Wolf}}]{perfetti2007ultrafast}%
  \BibitemOpen
  \bibfield  {author} {\bibinfo {author} {\bibfnamefont {L.}~\bibnamefont
  {Perfetti}}, \bibinfo {author} {\bibfnamefont {P.}~\bibnamefont {Loukakos}},
  \bibinfo {author} {\bibfnamefont {M.}~\bibnamefont {Lisowski}}, \bibinfo
  {author} {\bibfnamefont {U.}~\bibnamefont {Bovensiepen}}, \bibinfo {author}
  {\bibfnamefont {H.}~\bibnamefont {Eisaki}}, \ and\ \bibinfo {author}
  {\bibfnamefont {M.}~\bibnamefont {Wolf}},\ }\href@noop {} {\bibfield
  {journal} {\bibinfo  {journal} {Phys. Rev. Lett.}\ }\textbf {\bibinfo
  {volume} {99}},\ \bibinfo {pages} {197001} (\bibinfo {year}
  {2007})}\BibitemShut {NoStop}%
\bibitem [{\citenamefont {Tang}\ \emph
  {et~al.}(2020{\natexlab{a}})\citenamefont {Tang}, \citenamefont {Wang},
  \citenamefont {Duan}, \citenamefont {Yang}, \citenamefont {Huang},
  \citenamefont {Guo}, \citenamefont {Qian},\ and\ \citenamefont
  {Zhang}}]{tang2020non}%
  \BibitemOpen
  \bibfield  {author} {\bibinfo {author} {\bibfnamefont {T.}~\bibnamefont
  {Tang}}, \bibinfo {author} {\bibfnamefont {H.}~\bibnamefont {Wang}}, \bibinfo
  {author} {\bibfnamefont {S.}~\bibnamefont {Duan}}, \bibinfo {author}
  {\bibfnamefont {Y.}~\bibnamefont {Yang}}, \bibinfo {author} {\bibfnamefont
  {C.}~\bibnamefont {Huang}}, \bibinfo {author} {\bibfnamefont
  {Y.}~\bibnamefont {Guo}}, \bibinfo {author} {\bibfnamefont {D.}~\bibnamefont
  {Qian}}, \ and\ \bibinfo {author} {\bibfnamefont {W.}~\bibnamefont {Zhang}},\
  }\href@noop {} {\bibfield  {journal} {\bibinfo  {journal} {Phys. Rev. B}\
  }\textbf {\bibinfo {volume} {101}},\ \bibinfo {pages} {235148} (\bibinfo
  {year} {2020}{\natexlab{a}})}\BibitemShut {NoStop}%
\bibitem [{\citenamefont {Okazaki}\ \emph {et~al.}(2018)\citenamefont
  {Okazaki}, \citenamefont {Ogawa}, \citenamefont {Suzuki}, \citenamefont
  {Yamamoto}, \citenamefont {Someya}, \citenamefont {Michimae}, \citenamefont
  {Watanabe}, \citenamefont {Lu}, \citenamefont {Nohara}, \citenamefont
  {Takagi}, \citenamefont {Katayama}, \citenamefont {Sawa}, \citenamefont
  {Fujisawa}, \citenamefont {Kanai}, \citenamefont {Ishii}, \citenamefont
  {Itatani}, \citenamefont {Mizokawa},\ and\ \citenamefont
  {Shin}}]{okazaki2018photo}%
  \BibitemOpen
  \bibfield  {author} {\bibinfo {author} {\bibfnamefont {K.}~\bibnamefont
  {Okazaki}}, \bibinfo {author} {\bibfnamefont {Y.}~\bibnamefont {Ogawa}},
  \bibinfo {author} {\bibfnamefont {T.}~\bibnamefont {Suzuki}}, \bibinfo
  {author} {\bibfnamefont {T.}~\bibnamefont {Yamamoto}}, \bibinfo {author}
  {\bibfnamefont {T.}~\bibnamefont {Someya}}, \bibinfo {author} {\bibfnamefont
  {S.}~\bibnamefont {Michimae}}, \bibinfo {author} {\bibfnamefont
  {M.}~\bibnamefont {Watanabe}}, \bibinfo {author} {\bibfnamefont
  {Y.}~\bibnamefont {Lu}}, \bibinfo {author} {\bibfnamefont {M.}~\bibnamefont
  {Nohara}}, \bibinfo {author} {\bibfnamefont {H.}~\bibnamefont {Takagi}},
  \bibinfo {author} {\bibfnamefont {N.}~\bibnamefont {Katayama}}, \bibinfo
  {author} {\bibfnamefont {H.}~\bibnamefont {Sawa}}, \bibinfo {author}
  {\bibfnamefont {M.}~\bibnamefont {Fujisawa}}, \bibinfo {author}
  {\bibfnamefont {T.}~\bibnamefont {Kanai}}, \bibinfo {author} {\bibfnamefont
  {N.}~\bibnamefont {Ishii}}, \bibinfo {author} {\bibfnamefont
  {J.}~\bibnamefont {Itatani}}, \bibinfo {author} {\bibfnamefont
  {T.}~\bibnamefont {Mizokawa}}, \ and\ \bibinfo {author} {\bibfnamefont
  {S.}~\bibnamefont {Shin}},\ }\href@noop {} {\bibfield  {journal} {\bibinfo
  {journal} {Nat. Commun.}\ }\textbf {\bibinfo {volume} {9}},\ \bibinfo {pages}
  {4322} (\bibinfo {year} {2018})}\BibitemShut {NoStop}%
\bibitem [{\citenamefont {Schmitt}\ \emph {et~al.}(2008)\citenamefont
  {Schmitt}, \citenamefont {Kirchmann}, \citenamefont {Bovensiepen},
  \citenamefont {Moore}, \citenamefont {Rettig}, \citenamefont {Krenz},
  \citenamefont {Chu}, \citenamefont {Ru}, \citenamefont {Perfetti},
  \citenamefont {Lu}, \citenamefont {Wolf}, \citenamefont {Fisher},\ and\
  \citenamefont {Shen}}]{schmitt2008transient}%
  \BibitemOpen
  \bibfield  {author} {\bibinfo {author} {\bibfnamefont {F.}~\bibnamefont
  {Schmitt}}, \bibinfo {author} {\bibfnamefont {P.~S.}\ \bibnamefont
  {Kirchmann}}, \bibinfo {author} {\bibfnamefont {U.}~\bibnamefont
  {Bovensiepen}}, \bibinfo {author} {\bibfnamefont {R.~G.}\ \bibnamefont
  {Moore}}, \bibinfo {author} {\bibfnamefont {L.}~\bibnamefont {Rettig}},
  \bibinfo {author} {\bibfnamefont {M.}~\bibnamefont {Krenz}}, \bibinfo
  {author} {\bibfnamefont {J.-H.}\ \bibnamefont {Chu}}, \bibinfo {author}
  {\bibfnamefont {N.}~\bibnamefont {Ru}}, \bibinfo {author} {\bibfnamefont
  {L.}~\bibnamefont {Perfetti}}, \bibinfo {author} {\bibfnamefont {D.~H.}\
  \bibnamefont {Lu}}, \bibinfo {author} {\bibfnamefont {M.}~\bibnamefont
  {Wolf}}, \bibinfo {author} {\bibfnamefont {I.~R.}\ \bibnamefont {Fisher}}, \
  and\ \bibinfo {author} {\bibfnamefont {Z.-X.}\ \bibnamefont {Shen}},\
  }\href@noop {} {\bibfield  {journal} {\bibinfo  {journal} {Science}\ }\textbf
  {\bibinfo {volume} {321}},\ \bibinfo {pages} {1649} (\bibinfo {year}
  {2008})}\BibitemShut {NoStop}%
\bibitem [{\citenamefont {Petersen}\ \emph {et~al.}(2011)\citenamefont
  {Petersen}, \citenamefont {Kaiser}, \citenamefont {Dean}, \citenamefont
  {Simoncig}, \citenamefont {Liu}, \citenamefont {Cavalieri}, \citenamefont
  {Cacho}, \citenamefont {Turcu}, \citenamefont {Springate}, \citenamefont
  {Frassetto}, \citenamefont {Poletto}, \citenamefont {Dhesi}, \citenamefont
  {Berger},\ and\ \citenamefont {Cavalleri}}]{petersen2011clocking}%
  \BibitemOpen
  \bibfield  {author} {\bibinfo {author} {\bibfnamefont {J.~C.}\ \bibnamefont
  {Petersen}}, \bibinfo {author} {\bibfnamefont {S.}~\bibnamefont {Kaiser}},
  \bibinfo {author} {\bibfnamefont {N.}~\bibnamefont {Dean}}, \bibinfo {author}
  {\bibfnamefont {A.}~\bibnamefont {Simoncig}}, \bibinfo {author}
  {\bibfnamefont {H.~Y.}\ \bibnamefont {Liu}}, \bibinfo {author} {\bibfnamefont
  {A.~L.}\ \bibnamefont {Cavalieri}}, \bibinfo {author} {\bibfnamefont
  {C.}~\bibnamefont {Cacho}}, \bibinfo {author} {\bibfnamefont {I.~C.~E.}\
  \bibnamefont {Turcu}}, \bibinfo {author} {\bibfnamefont {E.}~\bibnamefont
  {Springate}}, \bibinfo {author} {\bibfnamefont {F.}~\bibnamefont
  {Frassetto}}, \bibinfo {author} {\bibfnamefont {L.}~\bibnamefont {Poletto}},
  \bibinfo {author} {\bibfnamefont {S.~S.}\ \bibnamefont {Dhesi}}, \bibinfo
  {author} {\bibfnamefont {H.}~\bibnamefont {Berger}}, \ and\ \bibinfo {author}
  {\bibfnamefont {A.}~\bibnamefont {Cavalleri}},\ }\href@noop {} {\bibfield
  {journal} {\bibinfo  {journal} {Phys. Rev. Lett.}\ }\textbf {\bibinfo
  {volume} {107}},\ \bibinfo {pages} {177402} (\bibinfo {year}
  {2011})}\BibitemShut {NoStop}%
\bibitem [{\citenamefont {Rohwer}\ \emph {et~al.}(2011)\citenamefont {Rohwer},
  \citenamefont {Hellmann}, \citenamefont {Wiesenmayer}, \citenamefont {Sohrt},
  \citenamefont {Stange}, \citenamefont {Slomski}, \citenamefont {Carr},
  \citenamefont {Liu}, \citenamefont {Avila}, \citenamefont {Kall{\"a}ne},
  \citenamefont {Mathias}, \citenamefont {Kipp}, \citenamefont {Rossnagel},\
  and\ \citenamefont {Bauer}}]{rohwer2011collapse}%
  \BibitemOpen
  \bibfield  {author} {\bibinfo {author} {\bibfnamefont {T.}~\bibnamefont
  {Rohwer}}, \bibinfo {author} {\bibfnamefont {S.}~\bibnamefont {Hellmann}},
  \bibinfo {author} {\bibfnamefont {M.}~\bibnamefont {Wiesenmayer}}, \bibinfo
  {author} {\bibfnamefont {C.}~\bibnamefont {Sohrt}}, \bibinfo {author}
  {\bibfnamefont {A.}~\bibnamefont {Stange}}, \bibinfo {author} {\bibfnamefont
  {B.}~\bibnamefont {Slomski}}, \bibinfo {author} {\bibfnamefont
  {A.}~\bibnamefont {Carr}}, \bibinfo {author} {\bibfnamefont {Y.}~\bibnamefont
  {Liu}}, \bibinfo {author} {\bibfnamefont {L.~M.}\ \bibnamefont {Avila}},
  \bibinfo {author} {\bibfnamefont {M.}~\bibnamefont {Kall{\"a}ne}}, \bibinfo
  {author} {\bibfnamefont {S.}~\bibnamefont {Mathias}}, \bibinfo {author}
  {\bibfnamefont {L.}~\bibnamefont {Kipp}}, \bibinfo {author} {\bibfnamefont
  {K.}~\bibnamefont {Rossnagel}}, \ and\ \bibinfo {author} {\bibfnamefont
  {M.}~\bibnamefont {Bauer}},\ }\href@noop {} {\bibfield  {journal} {\bibinfo
  {journal} {Nature}\ }\textbf {\bibinfo {volume} {471}},\ \bibinfo {pages}
  {490} (\bibinfo {year} {2011})}\BibitemShut {NoStop}%
\bibitem [{\citenamefont {Wang}\ \emph {et~al.}(2013)\citenamefont {Wang},
  \citenamefont {Steinberg}, \citenamefont {Jarillo-Herrero},\ and\
  \citenamefont {Gedik}}]{wang2013observation}%
  \BibitemOpen
  \bibfield  {author} {\bibinfo {author} {\bibfnamefont {Y.}~\bibnamefont
  {Wang}}, \bibinfo {author} {\bibfnamefont {H.}~\bibnamefont {Steinberg}},
  \bibinfo {author} {\bibfnamefont {P.}~\bibnamefont {Jarillo-Herrero}}, \ and\
  \bibinfo {author} {\bibfnamefont {N.}~\bibnamefont {Gedik}},\ }\href@noop {}
  {\bibfield  {journal} {\bibinfo  {journal} {Science}\ }\textbf {\bibinfo
  {volume} {342}},\ \bibinfo {pages} {453} (\bibinfo {year}
  {2013})}\BibitemShut {NoStop}%
\bibitem [{\citenamefont {Mahmood}\ \emph {et~al.}(2016)\citenamefont
  {Mahmood}, \citenamefont {Chan}, \citenamefont {Alpichshev}, \citenamefont
  {Gardner}, \citenamefont {Lee}, \citenamefont {Lee},\ and\ \citenamefont
  {Gedik}}]{mahmood2016selective}%
  \BibitemOpen
  \bibfield  {author} {\bibinfo {author} {\bibfnamefont {F.}~\bibnamefont
  {Mahmood}}, \bibinfo {author} {\bibfnamefont {C.-K.}\ \bibnamefont {Chan}},
  \bibinfo {author} {\bibfnamefont {Z.}~\bibnamefont {Alpichshev}}, \bibinfo
  {author} {\bibfnamefont {D.}~\bibnamefont {Gardner}}, \bibinfo {author}
  {\bibfnamefont {Y.}~\bibnamefont {Lee}}, \bibinfo {author} {\bibfnamefont
  {P.~A.}\ \bibnamefont {Lee}}, \ and\ \bibinfo {author} {\bibfnamefont
  {N.}~\bibnamefont {Gedik}},\ }\href@noop {} {\bibfield  {journal} {\bibinfo
  {journal} {Nat. Phys.}\ }\textbf {\bibinfo {volume} {12}},\ \bibinfo {pages}
  {306} (\bibinfo {year} {2016})}\BibitemShut {NoStop}%
\bibitem [{\citenamefont {Zhou}\ \emph {et~al.}(2023)\citenamefont {Zhou},
  \citenamefont {Bao}, \citenamefont {Fan}, \citenamefont {Zhou}, \citenamefont
  {Gao}, \citenamefont {Zhong}, \citenamefont {Lin}, \citenamefont {Liu},
  \citenamefont {Yu}, \citenamefont {Tang}, \citenamefont {Meng}, \citenamefont
  {Duan},\ and\ \citenamefont {Zhou}}]{zhou2023pseudospin}%
  \BibitemOpen
  \bibfield  {author} {\bibinfo {author} {\bibfnamefont {S.}~\bibnamefont
  {Zhou}}, \bibinfo {author} {\bibfnamefont {C.}~\bibnamefont {Bao}}, \bibinfo
  {author} {\bibfnamefont {B.}~\bibnamefont {Fan}}, \bibinfo {author}
  {\bibfnamefont {H.}~\bibnamefont {Zhou}}, \bibinfo {author} {\bibfnamefont
  {Q.}~\bibnamefont {Gao}}, \bibinfo {author} {\bibfnamefont {H.}~\bibnamefont
  {Zhong}}, \bibinfo {author} {\bibfnamefont {T.}~\bibnamefont {Lin}}, \bibinfo
  {author} {\bibfnamefont {H.}~\bibnamefont {Liu}}, \bibinfo {author}
  {\bibfnamefont {P.}~\bibnamefont {Yu}}, \bibinfo {author} {\bibfnamefont
  {P.}~\bibnamefont {Tang}}, \bibinfo {author} {\bibfnamefont {S.}~\bibnamefont
  {Meng}}, \bibinfo {author} {\bibfnamefont {W.}~\bibnamefont {Duan}}, \ and\
  \bibinfo {author} {\bibfnamefont {S.}~\bibnamefont {Zhou}},\ }\href@noop {}
  {\bibfield  {journal} {\bibinfo  {journal} {Nature}\ }\textbf {\bibinfo
  {volume} {614}},\ \bibinfo {pages} {75} (\bibinfo {year} {2023})}\BibitemShut
  {NoStop}%
\bibitem [{\citenamefont {Cavalieri}\ \emph {et~al.}(2007)\citenamefont
  {Cavalieri}, \citenamefont {M{\"u}ller}, \citenamefont {Uphues},
  \citenamefont {Yakovlev}, \citenamefont {Baltu{\v{s}}ka}, \citenamefont
  {Horvath}, \citenamefont {Schmidt}, \citenamefont {Bl{\"u}mel}, \citenamefont
  {Holzwarth}, \citenamefont {Hendel}, \citenamefont {Drescher}, \citenamefont
  {Kleineberg}, \citenamefont {Echenique}, \citenamefont {Kienberger},
  \citenamefont {Krausz},\ and\ \citenamefont
  {Heinzmann}}]{cavalieri2007attosecond}%
  \BibitemOpen
  \bibfield  {author} {\bibinfo {author} {\bibfnamefont {A.~L.}\ \bibnamefont
  {Cavalieri}}, \bibinfo {author} {\bibfnamefont {N.}~\bibnamefont
  {M{\"u}ller}}, \bibinfo {author} {\bibfnamefont {T.}~\bibnamefont {Uphues}},
  \bibinfo {author} {\bibfnamefont {V.~S.}\ \bibnamefont {Yakovlev}}, \bibinfo
  {author} {\bibfnamefont {A.}~\bibnamefont {Baltu{\v{s}}ka}}, \bibinfo
  {author} {\bibfnamefont {B.}~\bibnamefont {Horvath}}, \bibinfo {author}
  {\bibfnamefont {B.}~\bibnamefont {Schmidt}}, \bibinfo {author} {\bibfnamefont
  {L.}~\bibnamefont {Bl{\"u}mel}}, \bibinfo {author} {\bibfnamefont
  {R.}~\bibnamefont {Holzwarth}}, \bibinfo {author} {\bibfnamefont
  {S.}~\bibnamefont {Hendel}}, \bibinfo {author} {\bibfnamefont
  {M.}~\bibnamefont {Drescher}}, \bibinfo {author} {\bibfnamefont
  {U.}~\bibnamefont {Kleineberg}}, \bibinfo {author} {\bibfnamefont {P.~M.}\
  \bibnamefont {Echenique}}, \bibinfo {author} {\bibfnamefont {R.}~\bibnamefont
  {Kienberger}}, \bibinfo {author} {\bibfnamefont {F.}~\bibnamefont {Krausz}},
  \ and\ \bibinfo {author} {\bibfnamefont {U.}~\bibnamefont {Heinzmann}},\
  }\href@noop {} {\bibfield  {journal} {\bibinfo  {journal} {Nature}\ }\textbf
  {\bibinfo {volume} {449}},\ \bibinfo {pages} {1029} (\bibinfo {year}
  {2007})}\BibitemShut {NoStop}%
\bibitem [{\citenamefont {Tao}\ \emph {et~al.}(2016)\citenamefont {Tao},
  \citenamefont {Chen}, \citenamefont {Szilv{\'a}si}, \citenamefont {Keller},
  \citenamefont {Mavrikakis}, \citenamefont {Kapteyn},\ and\ \citenamefont
  {Murnane}}]{tao2016direct}%
  \BibitemOpen
  \bibfield  {author} {\bibinfo {author} {\bibfnamefont {Z.}~\bibnamefont
  {Tao}}, \bibinfo {author} {\bibfnamefont {C.}~\bibnamefont {Chen}}, \bibinfo
  {author} {\bibfnamefont {T.}~\bibnamefont {Szilv{\'a}si}}, \bibinfo {author}
  {\bibfnamefont {M.}~\bibnamefont {Keller}}, \bibinfo {author} {\bibfnamefont
  {M.}~\bibnamefont {Mavrikakis}}, \bibinfo {author} {\bibfnamefont
  {H.}~\bibnamefont {Kapteyn}}, \ and\ \bibinfo {author} {\bibfnamefont
  {M.}~\bibnamefont {Murnane}},\ }\href@noop {} {\bibfield  {journal} {\bibinfo
   {journal} {Science}\ }\textbf {\bibinfo {volume} {353}},\ \bibinfo {pages}
  {62} (\bibinfo {year} {2016})}\BibitemShut {NoStop}%
\bibitem [{\citenamefont {Yang}\ \emph {et~al.}(2019)\citenamefont {Yang},
  \citenamefont {Tang}, \citenamefont {Duan}, \citenamefont {Zhou},
  \citenamefont {Hao},\ and\ \citenamefont {Zhang}}]{yang2019time}%
  \BibitemOpen
  \bibfield  {author} {\bibinfo {author} {\bibfnamefont {Y.}~\bibnamefont
  {Yang}}, \bibinfo {author} {\bibfnamefont {T.}~\bibnamefont {Tang}}, \bibinfo
  {author} {\bibfnamefont {S.}~\bibnamefont {Duan}}, \bibinfo {author}
  {\bibfnamefont {C.}~\bibnamefont {Zhou}}, \bibinfo {author} {\bibfnamefont
  {D.}~\bibnamefont {Hao}}, \ and\ \bibinfo {author} {\bibfnamefont
  {W.}~\bibnamefont {Zhang}},\ }\href@noop {} {\bibfield  {journal} {\bibinfo
  {journal} {Rev. Sci. Instrum.}\ }\textbf {\bibinfo {volume} {90}},\ \bibinfo
  {pages} {063905} (\bibinfo {year} {2019})}\BibitemShut {NoStop}%
\bibitem [{\citenamefont {Bao}\ \emph {et~al.}(2022)\citenamefont {Bao},
  \citenamefont {Zhong}, \citenamefont {Zhou}, \citenamefont {Feng},
  \citenamefont {Wang},\ and\ \citenamefont {Zhou}}]{bao2022ultrafast}%
  \BibitemOpen
  \bibfield  {author} {\bibinfo {author} {\bibfnamefont {C.}~\bibnamefont
  {Bao}}, \bibinfo {author} {\bibfnamefont {H.}~\bibnamefont {Zhong}}, \bibinfo
  {author} {\bibfnamefont {S.}~\bibnamefont {Zhou}}, \bibinfo {author}
  {\bibfnamefont {R.}~\bibnamefont {Feng}}, \bibinfo {author} {\bibfnamefont
  {Y.}~\bibnamefont {Wang}}, \ and\ \bibinfo {author} {\bibfnamefont
  {S.}~\bibnamefont {Zhou}},\ }\href@noop {} {\bibfield  {journal} {\bibinfo
  {journal} {Rev. Sci. Instrum.}\ }\textbf {\bibinfo {volume} {93}},\ \bibinfo
  {pages} {013902} (\bibinfo {year} {2022})}\BibitemShut {NoStop}%
\bibitem [{\citenamefont {Ishida}\ \emph {et~al.}(2016)\citenamefont {Ishida},
  \citenamefont {Otsu}, \citenamefont {Ozawa}, \citenamefont {Yaji},
  \citenamefont {Tani}, \citenamefont {Shin},\ and\ \citenamefont
  {Kobayashi}}]{ishida2016high}%
  \BibitemOpen
  \bibfield  {author} {\bibinfo {author} {\bibfnamefont {Y.}~\bibnamefont
  {Ishida}}, \bibinfo {author} {\bibfnamefont {T.}~\bibnamefont {Otsu}},
  \bibinfo {author} {\bibfnamefont {A.}~\bibnamefont {Ozawa}}, \bibinfo
  {author} {\bibfnamefont {K.}~\bibnamefont {Yaji}}, \bibinfo {author}
  {\bibfnamefont {S.}~\bibnamefont {Tani}}, \bibinfo {author} {\bibfnamefont
  {S.}~\bibnamefont {Shin}}, \ and\ \bibinfo {author} {\bibfnamefont
  {Y.}~\bibnamefont {Kobayashi}},\ }\href@noop {} {\bibfield  {journal}
  {\bibinfo  {journal} {Rev. Sci. Instrum.}\ }\textbf {\bibinfo {volume}
  {87}},\ \bibinfo {pages} {123902} (\bibinfo {year} {2016})}\BibitemShut
  {NoStop}%
\bibitem [{\citenamefont {Chen}\ \emph {et~al.}(2023)\citenamefont {Chen},
  \citenamefont {Wang}, \citenamefont {Pan}, \citenamefont {Liu}, \citenamefont
  {Huang}, \citenamefont {Zhao}, \citenamefont {Yun}, \citenamefont {Qian},
  \citenamefont {Wei},\ and\ \citenamefont {Ding}}]{chen2023time}%
  \BibitemOpen
  \bibfield  {author} {\bibinfo {author} {\bibfnamefont {F.}~\bibnamefont
  {Chen}}, \bibinfo {author} {\bibfnamefont {J.}~\bibnamefont {Wang}}, \bibinfo
  {author} {\bibfnamefont {M.}~\bibnamefont {Pan}}, \bibinfo {author}
  {\bibfnamefont {J.}~\bibnamefont {Liu}}, \bibinfo {author} {\bibfnamefont
  {J.}~\bibnamefont {Huang}}, \bibinfo {author} {\bibfnamefont
  {K.}~\bibnamefont {Zhao}}, \bibinfo {author} {\bibfnamefont {C.}~\bibnamefont
  {Yun}}, \bibinfo {author} {\bibfnamefont {T.}~\bibnamefont {Qian}}, \bibinfo
  {author} {\bibfnamefont {Z.}~\bibnamefont {Wei}}, \ and\ \bibinfo {author}
  {\bibfnamefont {H.}~\bibnamefont {Ding}},\ }\href@noop {} {\bibfield
  {journal} {\bibinfo  {journal} {Rev. Sci. Instrum.}\ }\textbf {\bibinfo
  {volume} {94}},\ \bibinfo {pages} {043905} (\bibinfo {year}
  {2023})}\BibitemShut {NoStop}%
\bibitem [{\citenamefont {Sie}\ \emph {et~al.}(2019)\citenamefont {Sie},
  \citenamefont {Rohwer}, \citenamefont {Lee},\ and\ \citenamefont
  {Gedik}}]{sie2019time}%
  \BibitemOpen
  \bibfield  {author} {\bibinfo {author} {\bibfnamefont {E.~J.}\ \bibnamefont
  {Sie}}, \bibinfo {author} {\bibfnamefont {T.}~\bibnamefont {Rohwer}},
  \bibinfo {author} {\bibfnamefont {C.}~\bibnamefont {Lee}}, \ and\ \bibinfo
  {author} {\bibfnamefont {N.}~\bibnamefont {Gedik}},\ }\href@noop {}
  {\bibfield  {journal} {\bibinfo  {journal} {Nat. Commun.}\ }\textbf {\bibinfo
  {volume} {10}},\ \bibinfo {pages} {3535} (\bibinfo {year}
  {2019})}\BibitemShut {NoStop}%
\bibitem [{\citenamefont {Rohde}\ \emph {et~al.}(2016)\citenamefont {Rohde},
  \citenamefont {Hendel}, \citenamefont {Stange}, \citenamefont {Hanff},
  \citenamefont {Oloff}, \citenamefont {Yang}, \citenamefont {Rossnagel},\ and\
  \citenamefont {Bauer}}]{rohde2016time}%
  \BibitemOpen
  \bibfield  {author} {\bibinfo {author} {\bibfnamefont {G.}~\bibnamefont
  {Rohde}}, \bibinfo {author} {\bibfnamefont {A.}~\bibnamefont {Hendel}},
  \bibinfo {author} {\bibfnamefont {A.}~\bibnamefont {Stange}}, \bibinfo
  {author} {\bibfnamefont {K.}~\bibnamefont {Hanff}}, \bibinfo {author}
  {\bibfnamefont {L.-P.}\ \bibnamefont {Oloff}}, \bibinfo {author}
  {\bibfnamefont {L.}~\bibnamefont {Yang}}, \bibinfo {author} {\bibfnamefont
  {K.}~\bibnamefont {Rossnagel}}, \ and\ \bibinfo {author} {\bibfnamefont
  {M.}~\bibnamefont {Bauer}},\ }\href@noop {} {\bibfield  {journal} {\bibinfo
  {journal} {Rev. Sci. Instrum.}\ }\textbf {\bibinfo {volume} {87}},\ \bibinfo
  {pages} {103102} (\bibinfo {year} {2016})}\BibitemShut {NoStop}%
\bibitem [{\citenamefont {Puppin}\ \emph {et~al.}(2019)\citenamefont {Puppin},
  \citenamefont {Deng}, \citenamefont {Nicholson}, \citenamefont {Feldl},
  \citenamefont {Schr{\"o}ter}, \citenamefont {Vita}, \citenamefont
  {Kirchmann}, \citenamefont {Monney}, \citenamefont {Rettig}, \citenamefont
  {Wolf},\ and\ \citenamefont {Ernstorfer}}]{puppin2019time}%
  \BibitemOpen
  \bibfield  {author} {\bibinfo {author} {\bibfnamefont {M.}~\bibnamefont
  {Puppin}}, \bibinfo {author} {\bibfnamefont {Y.}~\bibnamefont {Deng}},
  \bibinfo {author} {\bibfnamefont {C.~W.}\ \bibnamefont {Nicholson}}, \bibinfo
  {author} {\bibfnamefont {J.}~\bibnamefont {Feldl}}, \bibinfo {author}
  {\bibfnamefont {N.}~\bibnamefont {Schr{\"o}ter}}, \bibinfo {author}
  {\bibfnamefont {H.}~\bibnamefont {Vita}}, \bibinfo {author} {\bibfnamefont
  {P.}~\bibnamefont {Kirchmann}}, \bibinfo {author} {\bibfnamefont
  {C.}~\bibnamefont {Monney}}, \bibinfo {author} {\bibfnamefont
  {L.}~\bibnamefont {Rettig}}, \bibinfo {author} {\bibfnamefont
  {M.}~\bibnamefont {Wolf}}, \ and\ \bibinfo {author} {\bibfnamefont
  {R.}~\bibnamefont {Ernstorfer}},\ }\href@noop {} {\bibfield  {journal}
  {\bibinfo  {journal} {Rev. Sci. Instrum.}\ }\textbf {\bibinfo {volume}
  {90}},\ \bibinfo {pages} {023104} (\bibinfo {year} {2019})}\BibitemShut
  {NoStop}%
\bibitem [{\citenamefont {Mills}\ \emph {et~al.}(2019)\citenamefont {Mills},
  \citenamefont {Zhdanovich}, \citenamefont {Na}, \citenamefont {Boschini},
  \citenamefont {Razzoli}, \citenamefont {Michiardi}, \citenamefont
  {Sheyerman}, \citenamefont {Schneider}, \citenamefont {Hammond},
  \citenamefont {Süss}, \citenamefont {Felser}, \citenamefont {Damascelli},\
  and\ \citenamefont {Jones}}]{mills2019cavity}%
  \BibitemOpen
  \bibfield  {author} {\bibinfo {author} {\bibfnamefont {A.~K.}\ \bibnamefont
  {Mills}}, \bibinfo {author} {\bibfnamefont {S.}~\bibnamefont {Zhdanovich}},
  \bibinfo {author} {\bibfnamefont {M.~X.}\ \bibnamefont {Na}}, \bibinfo
  {author} {\bibfnamefont {F.}~\bibnamefont {Boschini}}, \bibinfo {author}
  {\bibfnamefont {E.}~\bibnamefont {Razzoli}}, \bibinfo {author} {\bibfnamefont
  {M.}~\bibnamefont {Michiardi}}, \bibinfo {author} {\bibfnamefont
  {A.}~\bibnamefont {Sheyerman}}, \bibinfo {author} {\bibfnamefont
  {M.}~\bibnamefont {Schneider}}, \bibinfo {author} {\bibfnamefont {T.~J.}\
  \bibnamefont {Hammond}}, \bibinfo {author} {\bibfnamefont {V.}~\bibnamefont
  {Süss}}, \bibinfo {author} {\bibfnamefont {C.}~\bibnamefont {Felser}},
  \bibinfo {author} {\bibfnamefont {A.}~\bibnamefont {Damascelli}}, \ and\
  \bibinfo {author} {\bibfnamefont {D.~J.}\ \bibnamefont {Jones}},\ }\href@noop
  {} {\bibfield  {journal} {\bibinfo  {journal} {Rev. Sci. Instrum.}\ }\textbf
  {\bibinfo {volume} {90}},\ \bibinfo {pages} {083001} (\bibinfo {year}
  {2019})}\BibitemShut {NoStop}%
\bibitem [{\citenamefont {Wang}\ \emph {et~al.}(2023)\citenamefont {Wang},
  \citenamefont {Chen}, \citenamefont {Pan}, \citenamefont {Xu}, \citenamefont
  {Lv}, \citenamefont {Liu}, \citenamefont {Li}, \citenamefont {Fang},
  \citenamefont {Chen}, \citenamefont {Zhu} \emph {et~al.}}]{wang2023high}%
  \BibitemOpen
  \bibfield  {author} {\bibinfo {author} {\bibfnamefont {J.}~\bibnamefont
  {Wang}}, \bibinfo {author} {\bibfnamefont {F.}~\bibnamefont {Chen}}, \bibinfo
  {author} {\bibfnamefont {M.}~\bibnamefont {Pan}}, \bibinfo {author}
  {\bibfnamefont {S.}~\bibnamefont {Xu}}, \bibinfo {author} {\bibfnamefont
  {R.}~\bibnamefont {Lv}}, \bibinfo {author} {\bibfnamefont {J.}~\bibnamefont
  {Liu}}, \bibinfo {author} {\bibfnamefont {Y.}~\bibnamefont {Li}}, \bibinfo
  {author} {\bibfnamefont {S.}~\bibnamefont {Fang}}, \bibinfo {author}
  {\bibfnamefont {Y.}~\bibnamefont {Chen}}, \bibinfo {author} {\bibfnamefont
  {J.}~\bibnamefont {Zhu}},  \emph {et~al.},\ }\href@noop {} {\bibfield
  {journal} {\bibinfo  {journal} {Optics Express}\ }\textbf {\bibinfo {volume}
  {31}},\ \bibinfo {pages} {9854} (\bibinfo {year} {2023})}\BibitemShut
  {NoStop}%
\bibitem [{\citenamefont {Wu}\ \emph {et~al.}(1996)\citenamefont {Wu},
  \citenamefont {Tang}, \citenamefont {Ye},\ and\ \citenamefont
  {Chen}}]{wu1996linear}%
  \BibitemOpen
  \bibfield  {author} {\bibinfo {author} {\bibfnamefont {B.}~\bibnamefont
  {Wu}}, \bibinfo {author} {\bibfnamefont {D.}~\bibnamefont {Tang}}, \bibinfo
  {author} {\bibfnamefont {N.}~\bibnamefont {Ye}}, \ and\ \bibinfo {author}
  {\bibfnamefont {C.}~\bibnamefont {Chen}},\ }\href@noop {} {\bibfield
  {journal} {\bibinfo  {journal} {Opt. Mater.}\ }\textbf {\bibinfo {volume}
  {5}},\ \bibinfo {pages} {105} (\bibinfo {year} {1996})}\BibitemShut {NoStop}%
\bibitem [{\citenamefont {Chen}\ \emph {et~al.}(2009)\citenamefont {Chen},
  \citenamefont {Wang}, \citenamefont {Wang},\ and\ \citenamefont
  {Xu}}]{chen2009deep}%
  \BibitemOpen
  \bibfield  {author} {\bibinfo {author} {\bibfnamefont {C.}~\bibnamefont
  {Chen}}, \bibinfo {author} {\bibfnamefont {G.}~\bibnamefont {Wang}}, \bibinfo
  {author} {\bibfnamefont {X.}~\bibnamefont {Wang}}, \ and\ \bibinfo {author}
  {\bibfnamefont {Z.}~\bibnamefont {Xu}},\ }\href@noop {} {\bibfield  {journal}
  {\bibinfo  {journal} {Appl. Phys. B}\ }\textbf {\bibinfo {volume} {97}},\
  \bibinfo {pages} {9} (\bibinfo {year} {2009})}\BibitemShut {NoStop}%
\bibitem [{\citenamefont {Zhang}\ \emph {et~al.}(2009)\citenamefont {Zhang},
  \citenamefont {Liu}, \citenamefont {Qi}, \citenamefont {Dai}, \citenamefont
  {Fang},\ and\ \citenamefont {Zhang}}]{zhang2009topological}%
  \BibitemOpen
  \bibfield  {author} {\bibinfo {author} {\bibfnamefont {H.}~\bibnamefont
  {Zhang}}, \bibinfo {author} {\bibfnamefont {C.-X.}\ \bibnamefont {Liu}},
  \bibinfo {author} {\bibfnamefont {X.-L.}\ \bibnamefont {Qi}}, \bibinfo
  {author} {\bibfnamefont {X.}~\bibnamefont {Dai}}, \bibinfo {author}
  {\bibfnamefont {Z.}~\bibnamefont {Fang}}, \ and\ \bibinfo {author}
  {\bibfnamefont {S.-C.}\ \bibnamefont {Zhang}},\ }\href@noop {} {\bibfield
  {journal} {\bibinfo  {journal} {Nat. Phys}\ }\textbf {\bibinfo {volume}
  {5}},\ \bibinfo {pages} {438} (\bibinfo {year} {2009})}\BibitemShut {NoStop}%
\bibitem [{\citenamefont {Xia}\ \emph {et~al.}(2009)\citenamefont {Xia},
  \citenamefont {Qian}, \citenamefont {Hsieh}, \citenamefont {Wray},
  \citenamefont {Pal}, \citenamefont {Lin}, \citenamefont {Bansil},
  \citenamefont {Grauer}, \citenamefont {Hor}, \citenamefont {Cava},\ and\
  \citenamefont {Hasan}}]{xia2009observation}%
  \BibitemOpen
  \bibfield  {author} {\bibinfo {author} {\bibfnamefont {Y.}~\bibnamefont
  {Xia}}, \bibinfo {author} {\bibfnamefont {D.}~\bibnamefont {Qian}}, \bibinfo
  {author} {\bibfnamefont {D.}~\bibnamefont {Hsieh}}, \bibinfo {author}
  {\bibfnamefont {L.}~\bibnamefont {Wray}}, \bibinfo {author} {\bibfnamefont
  {A.}~\bibnamefont {Pal}}, \bibinfo {author} {\bibfnamefont {H.}~\bibnamefont
  {Lin}}, \bibinfo {author} {\bibfnamefont {A.}~\bibnamefont {Bansil}},
  \bibinfo {author} {\bibfnamefont {D.}~\bibnamefont {Grauer}}, \bibinfo
  {author} {\bibfnamefont {Y.~S.}\ \bibnamefont {Hor}}, \bibinfo {author}
  {\bibfnamefont {R.~J.}\ \bibnamefont {Cava}}, \ and\ \bibinfo {author}
  {\bibfnamefont {M.~Z.}\ \bibnamefont {Hasan}},\ }\href@noop {} {\bibfield
  {journal} {\bibinfo  {journal} {Nat. Phys}\ }\textbf {\bibinfo {volume}
  {5}},\ \bibinfo {pages} {398} (\bibinfo {year} {2009})}\BibitemShut {NoStop}%
\bibitem [{\citenamefont {Li}\ \emph {et~al.}(2019)\citenamefont {Li},
  \citenamefont {Gao}, \citenamefont {Duan}, \citenamefont {Xu}, \citenamefont
  {Zhu}, \citenamefont {Tian}, \citenamefont {Gao}, \citenamefont {Fan},
  \citenamefont {Rao}, \citenamefont {Huang}, \citenamefont {Li}, \citenamefont
  {Yan}, \citenamefont {Liu}, \citenamefont {Liu}, \citenamefont {Huang},
  \citenamefont {Li}, \citenamefont {Liu}, \citenamefont {Zhang}, \citenamefont
  {Zhang}, \citenamefont {Kondo}, \citenamefont {Shin}, \citenamefont {Lei},
  \citenamefont {Shi}, \citenamefont {Zhang}, \citenamefont {Weng},
  \citenamefont {Qian},\ and\ \citenamefont {Ding}}]{li2019dirac}%
  \BibitemOpen
  \bibfield  {author} {\bibinfo {author} {\bibfnamefont {H.}~\bibnamefont
  {Li}}, \bibinfo {author} {\bibfnamefont {S.-Y.}\ \bibnamefont {Gao}},
  \bibinfo {author} {\bibfnamefont {S.-F.}\ \bibnamefont {Duan}}, \bibinfo
  {author} {\bibfnamefont {Y.-F.}\ \bibnamefont {Xu}}, \bibinfo {author}
  {\bibfnamefont {K.-J.}\ \bibnamefont {Zhu}}, \bibinfo {author} {\bibfnamefont
  {S.-J.}\ \bibnamefont {Tian}}, \bibinfo {author} {\bibfnamefont {J.-C.}\
  \bibnamefont {Gao}}, \bibinfo {author} {\bibfnamefont {W.-H.}\ \bibnamefont
  {Fan}}, \bibinfo {author} {\bibfnamefont {Z.-C.}\ \bibnamefont {Rao}},
  \bibinfo {author} {\bibfnamefont {J.-R.}\ \bibnamefont {Huang}}, \bibinfo
  {author} {\bibfnamefont {J.-J.}\ \bibnamefont {Li}}, \bibinfo {author}
  {\bibfnamefont {D.-Y.}\ \bibnamefont {Yan}}, \bibinfo {author} {\bibfnamefont
  {Z.-T.}\ \bibnamefont {Liu}}, \bibinfo {author} {\bibfnamefont {W.-L.}\
  \bibnamefont {Liu}}, \bibinfo {author} {\bibfnamefont {Y.-B.}\ \bibnamefont
  {Huang}}, \bibinfo {author} {\bibfnamefont {Y.-L.}\ \bibnamefont {Li}},
  \bibinfo {author} {\bibfnamefont {Y.}~\bibnamefont {Liu}}, \bibinfo {author}
  {\bibfnamefont {G.-B.}\ \bibnamefont {Zhang}}, \bibinfo {author}
  {\bibfnamefont {P.}~\bibnamefont {Zhang}}, \bibinfo {author} {\bibfnamefont
  {T.}~\bibnamefont {Kondo}}, \bibinfo {author} {\bibfnamefont
  {S.}~\bibnamefont {Shin}}, \bibinfo {author} {\bibfnamefont {H.-C.}\
  \bibnamefont {Lei}}, \bibinfo {author} {\bibfnamefont {Y.-G.}\ \bibnamefont
  {Shi}}, \bibinfo {author} {\bibfnamefont {W.-T.}\ \bibnamefont {Zhang}},
  \bibinfo {author} {\bibfnamefont {H.-M.}\ \bibnamefont {Weng}}, \bibinfo
  {author} {\bibfnamefont {T.}~\bibnamefont {Qian}}, \ and\ \bibinfo {author}
  {\bibfnamefont {H.}~\bibnamefont {Ding}},\ }\href@noop {} {\bibfield
  {journal} {\bibinfo  {journal} {Phys. Rev. X.}\ }\textbf {\bibinfo {volume}
  {9}},\ \bibinfo {pages} {041039} (\bibinfo {year} {2019})}\BibitemShut
  {NoStop}%
\bibitem [{\citenamefont {Wakisaka}\ \emph {et~al.}(2009)\citenamefont
  {Wakisaka}, \citenamefont {Sudayama}, \citenamefont {Takubo}, \citenamefont
  {Mizokawa}, \citenamefont {Arita}, \citenamefont {Namatame}, \citenamefont
  {Taniguchi}, \citenamefont {Katayama}, \citenamefont {Nohara},\ and\
  \citenamefont {Takagi}}]{Wakisaka2009}%
  \BibitemOpen
  \bibfield  {author} {\bibinfo {author} {\bibfnamefont {Y.}~\bibnamefont
  {Wakisaka}}, \bibinfo {author} {\bibfnamefont {T.}~\bibnamefont {Sudayama}},
  \bibinfo {author} {\bibfnamefont {K.}~\bibnamefont {Takubo}}, \bibinfo
  {author} {\bibfnamefont {T.}~\bibnamefont {Mizokawa}}, \bibinfo {author}
  {\bibfnamefont {M.}~\bibnamefont {Arita}}, \bibinfo {author} {\bibfnamefont
  {H.}~\bibnamefont {Namatame}}, \bibinfo {author} {\bibfnamefont
  {M.}~\bibnamefont {Taniguchi}}, \bibinfo {author} {\bibfnamefont
  {N.}~\bibnamefont {Katayama}}, \bibinfo {author} {\bibfnamefont
  {M.}~\bibnamefont {Nohara}}, \ and\ \bibinfo {author} {\bibfnamefont
  {H.}~\bibnamefont {Takagi}},\ }\href@noop {} {\bibfield  {journal} {\bibinfo
  {journal} {Phys. Rev. Lett.}\ }\textbf {\bibinfo {volume} {103}},\ \bibinfo
  {pages} {026402} (\bibinfo {year} {2009})}\BibitemShut {NoStop}%
\bibitem [{\citenamefont {Mor}\ \emph {et~al.}(2017)\citenamefont {Mor},
  \citenamefont {Herzog}, \citenamefont {Golež}, \citenamefont {Werner},
  \citenamefont {Eckstein}, \citenamefont {Katayama}, \citenamefont {Nohara},
  \citenamefont {Takagi}, \citenamefont {Mizokawa}, \citenamefont {Monney},\
  and\ \citenamefont {Stähler}}]{Mor2017}%
  \BibitemOpen
  \bibfield  {author} {\bibinfo {author} {\bibfnamefont {S.}~\bibnamefont
  {Mor}}, \bibinfo {author} {\bibfnamefont {M.}~\bibnamefont {Herzog}},
  \bibinfo {author} {\bibfnamefont {D.}~\bibnamefont {Golež}}, \bibinfo
  {author} {\bibfnamefont {P.}~\bibnamefont {Werner}}, \bibinfo {author}
  {\bibfnamefont {M.}~\bibnamefont {Eckstein}}, \bibinfo {author}
  {\bibfnamefont {N.}~\bibnamefont {Katayama}}, \bibinfo {author}
  {\bibfnamefont {M.}~\bibnamefont {Nohara}}, \bibinfo {author} {\bibfnamefont
  {H.}~\bibnamefont {Takagi}}, \bibinfo {author} {\bibfnamefont
  {T.}~\bibnamefont {Mizokawa}}, \bibinfo {author} {\bibfnamefont
  {C.}~\bibnamefont {Monney}}, \ and\ \bibinfo {author} {\bibfnamefont
  {J.}~\bibnamefont {Stähler}},\ }\href@noop {} {\bibfield  {journal}
  {\bibinfo  {journal} {Phys. Rev. Lett.}\ }\textbf {\bibinfo {volume} {119}},\
  \bibinfo {pages} {086401} (\bibinfo {year} {2017})}\BibitemShut {NoStop}%
\bibitem [{\citenamefont {Fukutani}\ \emph {et~al.}(2021)\citenamefont
  {Fukutani}, \citenamefont {Stania}, \citenamefont {Kwon}, \citenamefont
  {Kim}, \citenamefont {Kong}, \citenamefont {Kim},\ and\ \citenamefont
  {Yeom}}]{Fukutani2021}%
  \BibitemOpen
  \bibfield  {author} {\bibinfo {author} {\bibfnamefont {K.}~\bibnamefont
  {Fukutani}}, \bibinfo {author} {\bibfnamefont {R.}~\bibnamefont {Stania}},
  \bibinfo {author} {\bibfnamefont {C.~I.}\ \bibnamefont {Kwon}}, \bibinfo
  {author} {\bibfnamefont {J.~S.}\ \bibnamefont {Kim}}, \bibinfo {author}
  {\bibfnamefont {K.~J.}\ \bibnamefont {Kong}}, \bibinfo {author}
  {\bibfnamefont {J.}~\bibnamefont {Kim}}, \ and\ \bibinfo {author}
  {\bibfnamefont {H.~W.}\ \bibnamefont {Yeom}},\ }\href@noop {} {\bibfield
  {journal} {\bibinfo  {journal} {Nat. Phys.}\ }\textbf {\bibinfo {volume}
  {17}},\ \bibinfo {pages} {1024} (\bibinfo {year} {2021})}\BibitemShut
  {NoStop}%
\bibitem [{\citenamefont {Lavenu}\ \emph {et~al.}(2018)\citenamefont {Lavenu},
  \citenamefont {Natile}, \citenamefont {Guichard}, \citenamefont {Zaouter},
  \citenamefont {Delen}, \citenamefont {Hanna}, \citenamefont {Mottay},\ and\
  \citenamefont {Georges}}]{lavenu2018nonlinear}%
  \BibitemOpen
  \bibfield  {author} {\bibinfo {author} {\bibfnamefont {L.}~\bibnamefont
  {Lavenu}}, \bibinfo {author} {\bibfnamefont {M.}~\bibnamefont {Natile}},
  \bibinfo {author} {\bibfnamefont {F.}~\bibnamefont {Guichard}}, \bibinfo
  {author} {\bibfnamefont {Y.}~\bibnamefont {Zaouter}}, \bibinfo {author}
  {\bibfnamefont {X.}~\bibnamefont {Delen}}, \bibinfo {author} {\bibfnamefont
  {M.}~\bibnamefont {Hanna}}, \bibinfo {author} {\bibfnamefont
  {E.}~\bibnamefont {Mottay}}, \ and\ \bibinfo {author} {\bibfnamefont
  {P.}~\bibnamefont {Georges}},\ }\href@noop {} {\bibfield  {journal} {\bibinfo
   {journal} {Opt. Lett.}\ }\textbf {\bibinfo {volume} {43}},\ \bibinfo {pages}
  {2252} (\bibinfo {year} {2018})}\BibitemShut {NoStop}%
\bibitem [{\citenamefont {Weitenberg}\ \emph {et~al.}(2017)\citenamefont
  {Weitenberg}, \citenamefont {Saule}, \citenamefont {Schulte},\ and\
  \citenamefont {Ru{\ss}b{\"u}ldt}}]{weitenberg2017nonlinear}%
  \BibitemOpen
  \bibfield  {author} {\bibinfo {author} {\bibfnamefont {J.}~\bibnamefont
  {Weitenberg}}, \bibinfo {author} {\bibfnamefont {T.}~\bibnamefont {Saule}},
  \bibinfo {author} {\bibfnamefont {J.}~\bibnamefont {Schulte}}, \ and\
  \bibinfo {author} {\bibfnamefont {P.}~\bibnamefont {Ru{\ss}b{\"u}ldt}},\
  }\href@noop {} {\bibfield  {journal} {\bibinfo  {journal} {IEEE J. Quantum
  Electron.}\ }\textbf {\bibinfo {volume} {53}},\ \bibinfo {pages} {1}
  (\bibinfo {year} {2017})}\BibitemShut {NoStop}%
\bibitem [{\citenamefont {Bao}\ \emph {et~al.}(2021)\citenamefont {Bao},
  \citenamefont {Luo}, \citenamefont {Zhang}, \citenamefont {Zhou},
  \citenamefont {Ren},\ and\ \citenamefont {Zhou}}]{bao2021full}%
  \BibitemOpen
  \bibfield  {author} {\bibinfo {author} {\bibfnamefont {C.}~\bibnamefont
  {Bao}}, \bibinfo {author} {\bibfnamefont {L.}~\bibnamefont {Luo}}, \bibinfo
  {author} {\bibfnamefont {H.}~\bibnamefont {Zhang}}, \bibinfo {author}
  {\bibfnamefont {S.}~\bibnamefont {Zhou}}, \bibinfo {author} {\bibfnamefont
  {Z.}~\bibnamefont {Ren}}, \ and\ \bibinfo {author} {\bibfnamefont
  {S.}~\bibnamefont {Zhou}},\ }\href@noop {} {\bibfield  {journal} {\bibinfo
  {journal} {Rev. Sci. Instrum.}\ }\textbf {\bibinfo {volume} {92}},\ \bibinfo
  {pages} {033904} (\bibinfo {year} {2021})}\BibitemShut {NoStop}%
\bibitem [{\citenamefont {Zhong}\ \emph {et~al.}(2023)\citenamefont {Zhong},
  \citenamefont {Liu}, \citenamefont {Wu}, \citenamefont {Guguchia},
  \citenamefont {Yin}, \citenamefont {Mine}, \citenamefont {Li}, \citenamefont
  {Najafzadeh}, \citenamefont {Das}, \citenamefont {Mielke~III}, \citenamefont
  {Khasanov}, \citenamefont {Luetkens}, \citenamefont {Suzuki}, \citenamefont
  {Liu}, \citenamefont {Han}, \citenamefont {Kondo}, \citenamefont {Hu},
  \citenamefont {Shin}, \citenamefont {Wang}, \citenamefont {Shi},
  \citenamefont {Yao},\ and\ \citenamefont {Okazaki}}]{zhong2023nodeless}%
  \BibitemOpen
  \bibfield  {author} {\bibinfo {author} {\bibfnamefont {Y.}~\bibnamefont
  {Zhong}}, \bibinfo {author} {\bibfnamefont {J.}~\bibnamefont {Liu}}, \bibinfo
  {author} {\bibfnamefont {X.}~\bibnamefont {Wu}}, \bibinfo {author}
  {\bibfnamefont {Z.}~\bibnamefont {Guguchia}}, \bibinfo {author}
  {\bibfnamefont {J.-X.}\ \bibnamefont {Yin}}, \bibinfo {author} {\bibfnamefont
  {A.}~\bibnamefont {Mine}}, \bibinfo {author} {\bibfnamefont {Y.}~\bibnamefont
  {Li}}, \bibinfo {author} {\bibfnamefont {S.}~\bibnamefont {Najafzadeh}},
  \bibinfo {author} {\bibfnamefont {D.}~\bibnamefont {Das}}, \bibinfo {author}
  {\bibfnamefont {C.}~\bibnamefont {Mielke~III}}, \bibinfo {author}
  {\bibfnamefont {R.}~\bibnamefont {Khasanov}}, \bibinfo {author}
  {\bibfnamefont {H.}~\bibnamefont {Luetkens}}, \bibinfo {author}
  {\bibfnamefont {T.}~\bibnamefont {Suzuki}}, \bibinfo {author} {\bibfnamefont
  {K.}~\bibnamefont {Liu}}, \bibinfo {author} {\bibfnamefont {X.}~\bibnamefont
  {Han}}, \bibinfo {author} {\bibfnamefont {T.}~\bibnamefont {Kondo}}, \bibinfo
  {author} {\bibfnamefont {J.}~\bibnamefont {Hu}}, \bibinfo {author}
  {\bibfnamefont {S.}~\bibnamefont {Shin}}, \bibinfo {author} {\bibfnamefont
  {Z.}~\bibnamefont {Wang}}, \bibinfo {author} {\bibfnamefont {X.}~\bibnamefont
  {Shi}}, \bibinfo {author} {\bibfnamefont {Y.}~\bibnamefont {Yao}}, \ and\
  \bibinfo {author} {\bibfnamefont {K.}~\bibnamefont {Okazaki}},\ }\href@noop
  {} {\bibfield  {journal} {\bibinfo  {journal} {Nature}\ ,\ \bibinfo {pages}
  {1}} (\bibinfo {year} {2023})}\BibitemShut {NoStop}%
\bibitem [{\citenamefont {Werdehausen}\ \emph {et~al.}(2018)\citenamefont
  {Werdehausen}, \citenamefont {Takayama}, \citenamefont {Höppner},
  \citenamefont {Albrecht}, \citenamefont {Rost}, \citenamefont {Lu},
  \citenamefont {Manske}, \citenamefont {Takagi},\ and\ \citenamefont
  {Kaiser}}]{Werdehausen2018}%
  \BibitemOpen
  \bibfield  {author} {\bibinfo {author} {\bibfnamefont {D.}~\bibnamefont
  {Werdehausen}}, \bibinfo {author} {\bibfnamefont {T.}~\bibnamefont
  {Takayama}}, \bibinfo {author} {\bibfnamefont {M.}~\bibnamefont {Höppner}},
  \bibinfo {author} {\bibfnamefont {G.}~\bibnamefont {Albrecht}}, \bibinfo
  {author} {\bibfnamefont {A.~W.}\ \bibnamefont {Rost}}, \bibinfo {author}
  {\bibfnamefont {Y.}~\bibnamefont {Lu}}, \bibinfo {author} {\bibfnamefont
  {D.}~\bibnamefont {Manske}}, \bibinfo {author} {\bibfnamefont
  {H.}~\bibnamefont {Takagi}}, \ and\ \bibinfo {author} {\bibfnamefont
  {S.}~\bibnamefont {Kaiser}},\ }\href@noop {} {\bibfield  {journal} {\bibinfo
  {journal} {Sci. Adv.}\ }\textbf {\bibinfo {volume} {4}},\ \bibinfo {pages}
  {eaap8652} (\bibinfo {year} {2018})}\BibitemShut {NoStop}%
\bibitem [{\citenamefont {Tang}\ \emph
  {et~al.}(2020{\natexlab{b}})\citenamefont {Tang}, \citenamefont {Wang},
  \citenamefont {Duan}, \citenamefont {Yang}, \citenamefont {Huang},
  \citenamefont {Guo}, \citenamefont {Qian},\ and\ \citenamefont
  {Zhang}}]{Tang2020}%
  \BibitemOpen
  \bibfield  {author} {\bibinfo {author} {\bibfnamefont {T.}~\bibnamefont
  {Tang}}, \bibinfo {author} {\bibfnamefont {H.}~\bibnamefont {Wang}}, \bibinfo
  {author} {\bibfnamefont {S.}~\bibnamefont {Duan}}, \bibinfo {author}
  {\bibfnamefont {Y.}~\bibnamefont {Yang}}, \bibinfo {author} {\bibfnamefont
  {C.}~\bibnamefont {Huang}}, \bibinfo {author} {\bibfnamefont
  {Y.}~\bibnamefont {Guo}}, \bibinfo {author} {\bibfnamefont {D.}~\bibnamefont
  {Qian}}, \ and\ \bibinfo {author} {\bibfnamefont {W.}~\bibnamefont {Zhang}},\
  }\href@noop {} {\bibfield  {journal} {\bibinfo  {journal} {Phys. Rev. B}\
  }\textbf {\bibinfo {volume} {101}},\ \bibinfo {pages} {235148} (\bibinfo
  {year} {2020}{\natexlab{b}})}\BibitemShut {NoStop}%
\end{thebibliography}%

\end{document}